\documentclass[nofootinbib,notitlepage,superscriptaddress,10pt,aps,prd,twocolumn]{revtex4-1}
\usepackage[utf8]{inputenc}
\usepackage{amsmath,mathtools}

\usepackage{tensor}
\usepackage{amssymb}
\usepackage{graphicx}
\usepackage{physics}
\usepackage{lipsum}  
\usepackage{bigints}
\usepackage{verbatim}
\usepackage{color}

\usepackage{caption}
\usepackage{subcaption}

\newcommand{\leri}[1]{\left(#1\right)}

\begin{document}
\title{Big-bounce and future time singularity resolution in Bianchi I: the projective invariant Nieh-Yan case}
\author{Flavio Bombacigno}
\email{flavio.bombacigno@ext.uv.es}
\affiliation{Departament de F\'{i}sica Teòrica and IFIC, Centro Mixto Universitat de València - CSIC, Universitat de València, Burjassot 46100, València, Spain}
\author{Simon Boudet}
\email{simon.boudet@unitn.it}
\affiliation{Dipartimento di Fisica, Universit\`{a} di Trento,\\Via Sommarive 14, I-38123 Povo (TN), Italy}
\affiliation{Trento Institute for Fundamental Physics and Applications (TIFPA)-INFN,\\Via Sommarive 14, I-38123 Povo (TN), Italy}
\author{Gonzalo J. Olmo}
\email{gonzalo.olmo@uv.es}
\affiliation{Departament de F\'{i}sica Teòrica and IFIC, Centro Mixto Universitat de València - CSIC, Universitat de València, Burjassot 46100, València, Spain}
\author{Giovanni Montani}
\email{giovanni.montani@enea.it}
\affiliation{Physics Department, ``Sapienza'' University of Rome,\\ P.le Aldo Moro 5, 00185 (Roma), Italy}
\affiliation{ENEA, Fusion and Nuclear Safety Department,\\ C. R. Frascati,
 	Via E. Fermi 45, 00044 Frascati (Roma), Italy}

\begin{abstract}
    We extend the notion of the Nieh-Yan invariant to generic metric-affine geometries, where both torsion and nonmetricity are taken into account. Notably, we show that the properties of projective invariance and topologicity can be independently accommodated by a suitable choice of the parameters featuring this new Nieh-Yan term. We then consider a special class of modified theories of gravity able to promote the Immirzi parameter to a dynamical scalar field coupled to the Nieh-Yan form, and we discuss in more detail the dynamics of the effective scalar tensor theory stemming from such a revised theoretical framework. We focus, in particular, on cosmological Bianchi I models and we derive classical solutions where the initial singularity is safely removed in favor of a big-bounce, which is ultimately driven by the non-minimal coupling with the Immirzi field. These solutions, moreover, turn out to be characterized by finite time singularities, but we show that such critical points do not spoil the geodesic completeness and wave regularity of these space-times.
\end{abstract}

\maketitle
\section{Introduction}
The theory of General Relativity (GR) \cite{WaldR.M.1984,MisnerC.W.2017} is based on the geometric interpretation of the gravitational field, described in terms of a metric tensor and a connection on a pseudo-Riemannian manifold. Both GR and many alternative theories of gravity rely on a metric formulation, in which the connection is completely determined by the metric tensor and its derivatives, resulting in the Levi-Civita connection, which is both symmetric and metric compatible. Geometric theories of gravity can also be formulated following the metric-affine paradigm, according to which the metric tensor and the connection are considered as independent variables. In this scheme, symmetry and metric compatibility of the connection are not a priori assumed, allowing for the presence of non vanishing torsion and nonmetricity tensors, respectively. Well known examples of metric-affine theories are Ricci based gravity \cite{Afonso:2018bpv,Afonso:2018hyj} (which encloses a large variety of sub cases, as for instance Palatini $f(R)$ theory \cite{Olmo2011}, quadratic gravity \cite{Lobo:2013adx}, and Born-Infeld-type models \cite{BeltranJimenez:2017doy}), general teleparallel models \cite{BELTRANJIMENEZ2020135422}, generalized hybrid metric-Palatini gravity \cite{Rosa:2021lhc,Harko:2011nh,Capozziello:2015lza,Bombacigno:2019did,Bronnikov:2019ugl} and metric-affine extension of higher order theories \cite{Borunda:2008kf,Janssen:2019doc,Janssen:2019uao,Percacci:2019hxn}.
\\ The connection plays a fundamental role also in one of the current attempts to quantize gravity, that is loop quantum gravity (LQG) \cite{Rovelli2004,Thiemann2007}, where the gravitational interaction is reformulated in terms of a gauge $SU(2)$ connection (Ashtekar-Barbero-Immirzi connection) and its conjugate momentum (densitized triad) \cite{Ashtekar1986,Ashtekar1987,Ashtekar1989,Ashtekar1992}. This representation, indeed, is usually derived by including an additional contribution to the first order (Palatini) action of GR, namely the Holst term \cite{Holst1996}, which results eventually vanishing when the equations of motion for the connection are satisfied (on half-shell). This guarantees the classical dynamics be preserved, and a proper set of smeared variables suitable for quantization is introduced \cite{Rovelli:1987df,Rovelli:1995ac}. We are mainly interested, however, in an equivalent formulation of LQG, which relies on the use of the Nieh-Yan (NY) topological invariant \cite{Nieh1982,Nieh2007} in place of the Holst term \cite{Date2009}. The NY invariant, initially discovered in the context of Riemann-Cartan theory, goes beyond the on half-shell vanishing character of the Holst term because of its main property, i.e. topologicity: it simply reduces to a boundary term without affecting the field equations at all. Now, these additional terms are driven by the so called Immirzi parameter $\beta$ \cite{Immirzi:1996di,Immirzi1997}, which is used in defining the Ashtekar variables and is related to a quantization ambiguity \cite{Rovelli:1997na}. Attempts to address this issue led to the proposal of considering the Immirzi parameter as a new fundamental field \cite{Taveras2008,Calcagni2009,Bombacigno2016}, an idea that has been later developed within more general modified gravity models \cite{Veraguth2017,Wang2018,Bombacigno2019,Bombacigno2018,Wang2020,Iosifidis:2020dck,BOMBACIGNO2021115281,Langvik:2020nrs}, revealing interesting phenomenology, such as bouncing solutions in isotropic cosmological models \cite{Taveras2008,Bombacigno2016,BombacignoFlavioandMontani2019} or hairy black hole solutions \cite{PhysRevD.103.084034}, together with implications at a more fundamental level regarding the strong CP problem \cite{PhysRevD.81.125015,PhysRevD.91.085017}, the chiral anomaly \cite{Mercuri2009a} and the implementation of Ashtekar variables \cite{Bombacigno2016,BOMBACIGNO2021115281}. The promotion of such constant parameter to a dynamical field is usually pursued ``by hand'', substituting $\beta\rightarrow \beta(x)$ in the Lagrangian and possibly adding a potential term $V(\beta)$. 
\\ More recently, there has been interest in the NY term also in the context of teleparallel gravity, where it was considered in the formulation of parity violating extensions of teleparallel models \cite{Li_2020} (see also \cite{Iosifidis:2020dck}, where the Holst term is taken into account), and in the field of condensed matter physics \cite{PhysRevResearch.2.033269,Nissinen2019,PhysRevB.101.125201,liu2021phonon}.
Beside topologicity, another important property featuring metric affine gravity is projective invariance \cite{Afonso:2017bxr,Iosifidis:2019fsh}, which has been recently shown to be of crucial importance for the dynamical stability of metric-affine theories \cite{BeltranJimenez2019}. This aspect demands special caution in the formulation of metric affine models featuring additional degrees of freedom, whose pathological nature may be determined by the presence of ghost-like instabilities. In this regard, we want to stress that, while the Holst term is projective invariant, the NY term breaks this symmetry, a feature usually neglected in literature. Therefore, in order to properly account for the projective symmetry in general metric-affine NY models, a revision of previous formulations seems necessary. Moreover, when considering a completely general metric-affine setup also the topological character of the NY term is lost, since it holds only for vanishing nonmetricity.
\\The approach followed in this note relies on the choice of including these features from the very beginning in the action, without imposing any restriction on the affine sector. We do this by proposing a generalization of the NY term to metric-affine geometries with arbitrary torsion and nonmetricity. We include two parameters in its definition, allowing to restore topologicity and projective invariance independently, keeping track of these two features separately. In particular, for appropriate values of the parameters, one can have projective invariance without topologicity, while the former is automatically implied by the latter. We then consider an action defined by a general function of two arguments, the Ricci scalar (built from the independent connection) and the generalized NY term, and perform the transformation to the Jordan frame. Here, we retain two additional scalar degrees of freedom and we propose to identify one of them with the Immirzi field. In this way we are able to induce a dynamical character for the Immirzi field and to include its own potential in a more natural way, without the need of introducing these features by hand in the action. Then, we study the effective scalar-tensor theory stemming from this model and we compare our results to previous treatments \cite{Calcagni2009,Mercuri2006,Mercuri2008,Mercuri2009,Mercuri2009a,Bombacigno2016,Bombacigno2018,Bombacigno2019} where metricity was a priori postulated and the role of projective symmetry neglected. We are able to reproduce such results for the appropriate values of the parameters featuring the generalized NY term and imposing the vanishing of nonmetricity via a Lagrange multiplier in the action. The fact that the usual Einstein-Cartan NY invariant and related models are properly recovered in this way, supports the correctness of our expression for the newly defined NY term, in favour of other possible generalizations preserving projective invariance and topologicity. Moreover, in the comparison with previous models, a further outcome stands out: despite the violation of projective symmetry in the action due to the choice of the parameters, projective invariance is somehow recovered on-shell, ensuring the absence of unstable modes, in contrast to \cite{BeltranJimenez2019}.
\\ 
After the formal discussion, our focus is put on the implementation of the constructed theory into a cosmological arena. In particular, we investigate the dynamics of a Bianchi I model (having zero spatial curvature and three distinct cosmic scale factors, each for each space direction) \cite{Montani:2011zz,Corichi:2009pp,Kamenshchik:2017ojc}, limiting our attention to the case in which the two basic parameters of the underlying Lagrangian are equal to unity. Such a restriction corresponds to dealing with a topological NY term and allows us to construct a semi-analytical solution for the considered cosmological model. For concreteness, we consider a quadratic  correction (recall that the theory relies on a Palatini approach) because this choice is natural in the spirit of extending the GR Lagrangian to the $f(R)$ domain and considering a Taylor series expansion.\\ 
The main result provided by our semi-analytical study of the Bianchi I model consists on the emergence of a classical bouncing cosmology (see also \cite{Montani:2021yom,Montani:2018bxv,Cianfrani:2012gv,Moriconi:2017bvs,Giovannetti:2021bqh,BombacignoFlavioandMontani2019,Barragan:2010qb,Barragan:2009sq}) for negative values of the parameter controlling the quadratic correction to the Ricci scalar. However, the interest for the present case is due to the presence of three degrees of freedom, each of them contributing, with its own specific behavior, to the universe volume dynamics. Indeed, while the universe volume naturally follows a bouncing evolution, characterized by a minimum value and a symmetric behavior before and after it, the evolution of the scale factors can introduce other features in the cosmological scenario. In particular, in the presence of matter (we include the contribution of an incoherent dust, mimicking the matter universe component and a radiation-like perfect fluid, corresponding to the primordial thermal bath energy-momentum), the value of such a parameter can drastically affect the dynamics after the bounce. Indeed, as long as pertaining to a specific range, scale factors suffer of singular points, where some of them may diverge while others approach zero at a given instant of time. We show that this cosmological picture is non-viable, because photon trajectories are dramatically affected, preventing their propagation forward, and scalar perturbations are not bounded. By contrast, in vacuum or when the quadratic correction parameter stays below a critical value, we recover a cosmological setting featured by instants in which the time derivative of the volume diverges (finite time singularity, see \cite{Nojiri:2005sx,Odintsov:2018uaw}). These divergences, however, do not preclude a reliable cosmological evolution. In fact, in this scenario the scale factors remain always finite and non-vanishing during their evolution and, in the presence of matter, their asymptotic behavior provides an isotropic late universe. Furthermore, we show that both photon paths and scalar perturbations have a regular behavior and phenomenology. In other words, neither the primordial black body radiation, i.e. the cosmic microwave background, nor its spectrum of scalar perturbation seem to be affected in a critical manner.\\ 
The paper is structured as follows. In Sec.~\ref{section 2} we briefly summarize the formalism of metric-affine gravity. In Sec.~\ref{section 3} the generalized NY invariant is presented and its properties are discussed. Sections \ref{section 4} and \ref{secIV} are devoted to the analysis of the gravitational model considered in this work, including the comparison of the results with previous treatments and the equivalence with DHOST theories. In Sec.~\ref{section 6} we present the cosmological solutions, whose properties are discussed in Sec.~\ref{geodesiccompleteness}. Eventually, conclusions are drawn in Sec.~\ref{conclusions}.
\\Spacetime signature is chosen mostly plus, i.e. $(-,+,+,+)$ and indices symmetrization and antisymmetrization is defined as $A_{(\mu\nu)}=(A_{\mu\nu}+A_{\nu\mu})/2$ and $A_{[\mu\nu]}=(A_{\mu\nu}-A_{\nu\mu})/2$, respectively.

\section{Formalism and notation}\label{section 2}
\noindent In this section we briefly discuss the notation we will adopt throughout the paper, in order to make the presentation as plain as possible. Since we consider metric-affine theories of gravity, where the connection $\Gamma\indices{^\rho_{\mu\nu}}$ is assumed to be an independent degree of freedom with respect to the metric field $g_{\mu\nu}$, we include in the analysis torsion and nonmetricity tensors, defined by:
\begin{equation}
    \begin{split}
        &T\indices{^\rho_{\mu\nu}}\equiv\Gamma\indices{^\rho_{\mu\nu}}-\Gamma\indices{^\rho_{\nu\mu}}, \\
        &Q\indices{_{\rho\mu\nu}}\equiv-\nabla_\rho g_{\mu\nu},
    \end{split}
\end{equation}
where we introduced the covariant derivative operation, denoted by $\nabla_\mu$ and acting as
\begin{equation}
    \nabla_\mu A\indices{^\rho_\sigma}=\partial_\mu A\indices{^\rho_\sigma}+\Gamma\indices{^\rho_{\lambda\mu}}A\indices{^\lambda_\sigma}-\Gamma\indices{^\lambda_{\sigma\mu}}A\indices{^\rho_\lambda}.
\end{equation}
Spacetime curvature is then encoded in the Riemann tensor, given by
\begin{equation}
    R\indices{^\rho_{\mu\sigma\nu}}=\partial_\sigma\Gamma\indices{^\rho_{\mu\nu}}-\partial_\nu\Gamma\indices{^\rho_{\mu\sigma}}+\Gamma\indices{^\rho_{\tau\sigma}}\Gamma\indices{^\tau_{\mu\nu}}-\Gamma\indices{^\rho_{\tau\nu}}\Gamma\indices{^\tau_{\mu\sigma}},
\end{equation}
and Ricci tensor and Ricci scalar are obtained from
\begin{equation}
    R_{\mu\nu}=R\indices{^\rho_{\mu\rho\nu}},\qquad R=g^{\mu\nu}R_{\mu\nu}.
\end{equation}
We note that when torsion and nonmetricity are taken into account, the Riemann tensor is skew-symmetric only in its last two indices and a further contraction, the so called homothetic curvature, can be built:
\begin{equation}
    \hat{R}_{\mu\nu}=R\indices{^\rho_{\rho\mu\nu}}.
\end{equation}
Now, it is useful to decompose torsion and nonmetricity in their irreducible parts. Concerning torsion, these are the trace vector
\begin{equation}
T_{\mu} \equiv T \indices{^{\nu}_{\mu\nu}},
\end{equation}
the pseudotrace axial vector
\begin{equation}
S_{\mu} \equiv \varepsilon_{\mu\nu\rho\sigma}T^{\nu\rho\sigma}
\end{equation}
and the antisymmetric tensor $q_{\rho\mu\nu}=-q_{\rho\nu\mu}$ satisfying
\begin{equation}
\varepsilon^{\mu\nu\rho\sigma} q_{\nu\rho\sigma} = 0, \qquad q\indices{^{\mu}_{\nu\mu}} = 0,
\end{equation}
which allow us to write the torsion tensor as
\begin{equation}
T_{\mu\nu\rho} = \dfrac{1}{3}\left(T_{\nu}g_{\mu\rho}-T_{\rho}g_{\mu\nu}\right) +\dfrac{1}{6} \varepsilon_{\mu\nu\rho\sigma}S^{\sigma} + q_{\mu\nu\rho}.
\label{torsion decomposition}
\end{equation}
Regarding the nonmetricity, instead, it can be split as
\begin{equation}
    Q_{\rho\mu\nu}=\frac{5Q_\rho-2P_\rho}{18}g_{\mu\nu}-\frac{Q_{(\mu}g_{\nu)\rho}-4P_{(\mu}g_{\nu)\rho}}{9}+\Omega_{\rho\mu\nu},
    \label{non metricity decomposition}
\end{equation}
where $Q_\rho=Q\indices{_\rho^\mu_\mu}$ is the Weyl vector, $\;P_\rho=Q\indices{^\mu_{\mu\rho}}=Q\indices{^\mu_{\rho\mu}}$ is the other independent trace and $\Omega\indices{_\rho^\mu_\mu}=\Omega\indices{^\mu_{\rho\mu}}=\Omega\indices{^\mu_{\mu\rho}}=0$ is the traceless part. Then, it is possible to rewrite the connection as
\begin{equation}
    \Gamma\indices{^\rho_{\mu\nu}}=\bar{\Gamma}\indices{^\rho_{\mu\nu}}+N\indices{^\rho_{\mu\nu}}=\bar{\Gamma}\indices{^\rho_{\mu\nu}}+K\indices{^\rho_{\mu\nu}}+D\indices{^\rho_{\mu\nu}},
    \label{decomposition connection}
\end{equation}
where we introduced the contorsion and disformal tensors
\begin{align}
    &K\indices{^\rho_{\mu\nu}}=-K\indices{_\mu^\rho_\nu}=\frac{1}{2}\leri{T\indices{^\rho_{\mu\nu}}-T\indices{_\mu^\rho_\nu}-T\indices{_\nu^\rho_\mu}},\\
    &D\indices{^\rho_{\mu\nu}}=D\indices{^\rho_{\nu\mu}}=\frac{1}{2}\leri{Q\indices{_{\mu\nu}^\rho}+Q\indices{_{\nu\mu}^\rho}-Q\indices{^\rho_{\mu\nu}}},
\end{align}
and we denoted by $\bar{\Gamma}\indices{^\rho_{\mu\nu}}$ the Levi Civita connection for the metric $g_{\mu\nu}$. We observe that the symmetric and the skew-symmetric part of the connection result in
\begin{align}
    &\Gamma\indices{^\rho_{(\mu\nu)}}=\bar{\Gamma}\indices{^\rho_{\mu\nu}}+K\indices{^\rho_{(\mu\nu)}}+D\indices{^\rho_{(\mu\nu)}},\\
    &\Gamma\indices{^\rho_{[\mu\nu]}}=K\indices{^\rho_{[\mu\nu]}}.
\end{align}
Finally, in terms of the distorsion tensor $N\indices{^\rho_{\mu\nu}}$, we can rewrite the Riemann tensor as
\begin{equation}
    R\indices{^\rho_{\mu\sigma\nu}}=\bar{R}\indices{^\rho_{\mu\sigma\nu}}+ 2 \bar{\nabla}_{[\sigma} N\indices{^\rho_{\mu|\nu]}} +2N\indices{^\rho_{\lambda[\sigma}}N\indices{^\lambda_{\mu|\nu]}},
    \label{decomposition riemann}
    \end{equation}
where bar quantities are built out of the Levi-Civita connection. 
\\ \noindent Now, let us introduce the notion of projective transformation acting on the connection, i.e.
\begin{equation}
    \tilde{\Gamma}\indices{^\rho_{\mu\nu}}=\Gamma\indices{^\rho_{\mu\nu}}+\delta\indices{^\rho_\mu}\xi_\nu,
    \label{projective}
\end{equation}
where $\xi_\mu$ represents an unspecified one-form degree of freedom, which implies the following transformation rules for torsion and nonmetricity: 
\begin{align}
    &\tilde{T}_{\lambda\mu\nu}=T_{\lambda\mu\nu}+g_{\lambda\mu}\xi_\nu-g_{\lambda\nu}\xi_\mu,
    \label{proj torsion}
    \\
    & \tilde{Q}_{\mu\nu\lambda}=Q_{\mu\nu\lambda}+2g_{\nu\lambda}\xi_\mu,
    \label{proj non metricity}
\end{align}
or in terms of their vector components:
\begin{align}
    &\tilde{T}^\rho= T^\rho-3\xi^\rho,
    \label{transformation components projective t}\\
    &\tilde{S}^\rho= S^\rho,\\
    &\tilde{Q}^\rho= Q^\rho+8\xi^\rho,\\
    &\tilde{P}^\rho= P^\rho+2\xi^\rho,
    \label{transformation components projective p}
\end{align}
while $q_{\mu\nu\rho}$ and $\Omega_{\mu\nu\rho}$ are left unchanged. Under \eqref{projective} the Riemann tensor transforms as
\begin{equation}
    \tilde{R}\indices{^\rho_{\mu\sigma\nu}}=R\indices{^\rho_{\mu\sigma\nu}}+\delta\indices{^\rho_\mu}(\partial_\sigma\xi_\nu-\partial_\nu\xi_\sigma)
    \label{projective transformation Riemann}
\end{equation}
and we see that only the symmetric part of the Ricci tensor remains unaffected, i.e.
\begin{equation}
    \tilde{R}_{\mu\nu}=R_{\mu\nu}+\partial_\mu\xi_\nu-\partial_\nu\xi_\mu,
\end{equation}
implying the invariance of the Ricci scalar $\tilde{R}=R$.

\section{The role of nonmetricity in the Nieh-Yan term}\label{section 3}
\noindent The starting point of our discussion is the observation that in the presence of nonmetricity the Nieh-Yan term \cite{Nieh1982,Nieh2007}, 
\begin{align}
    NY &\equiv\frac{1}{2}\varepsilon^{\mu\nu\rho\sigma}\leri{\frac{1}{2}T\indices{^\lambda_{\mu\nu}}T_{\lambda\rho\sigma}-R_{\mu\nu\rho\sigma}} \ ,
    \label{NY0}
\end{align}
is spoilt of its topological character. This can be seen  by taking into account \eqref{decomposition connection} and \eqref{decomposition riemann},  which lead to  
\begin{align}
    NY &=-\frac{1}{2}\bar{\nabla}\cdot S-\frac{1}{2}\varepsilon^{\mu\nu\rho\sigma}T\indices{^\lambda_{\mu\nu}}Q\indices{_{\rho\sigma\lambda}},
    \label{NY}
\end{align}
where we used
\begin{equation}
    \varepsilon^{\mu\nu\rho\sigma}R_{\mu\nu\rho\sigma}=\bar{\nabla}\cdot S+\frac{1}{2}\varepsilon^{\mu\nu\rho\sigma}T\indices{^\lambda_{\mu\nu}}(T_{\lambda\rho\sigma}+2Q\indices{_{\rho\sigma\lambda}}).
\end{equation}
It is therefore clear that when $Q_{\rho\mu\nu}\neq 0$ the Nieh-Yan term cannot be simply expressed as the divergence of a vector, and the appearance of nonmetricity explicitly breaks up topologicity. Even if in literature this feature has been always neglected by simply disregarding nonmetricity from the very beginning (see \cite{Calcagni2009,Mercuri2006,Mercuri2008,Mercuri2009,Mercuri2009a,Bombacigno2016,Bombacigno2018,Bombacigno2019}), when we are interested in a proper metric-affine generalization of LQG-inspired actions it seems sensible to look for extensions  of \eqref{NY} able to recover such a property. Moreover, because of the torsion tensor transformation rule \eqref{proj torsion}, we can easily verify that \eqref{NY} is also not invariant under projective transformations, i.e.
\begin{equation}
    \frac{1}{4}\varepsilon^{\mu\nu\rho\sigma}\tilde{T}\indices{^\lambda_{\mu\nu}}\tilde{T}_{\lambda\rho\sigma}-\frac{1}{4}\varepsilon^{\mu\nu\rho\sigma}T\indices{^\lambda_{\mu\nu}}T_{\lambda\rho\sigma}=-S_\mu\xi^\mu,
    \label{projective rule TT}
\end{equation}
and, in this respect, it has been recently suggested that projective breaking terms in the Lagrangian could be associated to dynamical instabilities\footnote{In particular, it has been outlined as the choice of neglecting torsion could offer a viable mechanism for restoring stability conditions.}, when higher order curvature terms are considered \cite{BeltranJimenez2019}. Now, by looking at \eqref{NY}, we point out that a newly topological Nieh-Yan term can be recovered by simply setting
\begin{equation}
    NY^{*}\equiv NY+\frac{1}{2}\varepsilon^{\mu\nu\rho\sigma}T\indices{^\lambda_{\mu\nu}}Q\indices{_{\rho\sigma\lambda}},
\end{equation}
which can be rewritten also as
\begin{equation}
    NY^{*}=\frac{1}{2}\varepsilon^{\mu\nu\rho\sigma}\leri{\frac{1}{2}T\indices{^\lambda_{\mu\nu}}(T_{\lambda\rho\sigma}+2Q\indices{_{\rho\sigma\lambda}})-R_{\mu\nu\rho\sigma}}.
    \label{NY top}
\end{equation}
We note that projective invariance is now enclosed as well, since
\begin{equation}
        \frac{1}{2}\varepsilon^{\mu\nu\rho\sigma}\tilde{T}\indices{^\lambda_{\mu\nu}}\tilde{Q}\indices{_{\rho\sigma\lambda}}-\frac{1}{2}\varepsilon^{\mu\nu\rho\sigma}T\indices{^\lambda_{\mu\nu}}Q\indices{_{\rho\sigma\lambda}}=+S_\mu\xi^\mu,
\end{equation}
which exactly cancels out \eqref{projective rule TT}. We stress, however, that projective invariance is not strictly related to topologicity, and suitable generalizations of \eqref{NY top} breaking up only with the latter can be actually formulated. Let us consider, for instance, the following modified Nieh-Yan term
\begin{equation}
    NY_{gen}\equiv\frac{1}{2}\varepsilon^{\mu\nu\rho\sigma}\leri{\frac{\lambda_1}{2}T\indices{^\lambda_{\mu\nu}}T_{\lambda\rho\sigma}+\lambda_2\, T\indices{^\lambda_{\mu\nu}}Q\indices{_{\rho\sigma\lambda}}-R_{\mu\nu\rho\sigma}},
    \label{NY general}
\end{equation}
where we introduced the real parameters $\lambda_1,\,\lambda_2$. In this case the term \eqref{NY general} transforms as
\begin{equation}
    NY_{gen}\rightarrow NY_{gen}-(\lambda_1-\lambda_2)\xi^\mu S_\mu,
\end{equation}
so that by setting $\lambda_1=\lambda_2$ we can recover again projective invariance, despite topologicity being in general violated if $\lambda_1=\lambda_2\neq 1$:
\begin{equation}
\begin{split}
     NY_{gen}=&-\frac{1}{2}\bar{\nabla}\cdot S+\\
     &+\frac{ (\lambda_1-1)}{4}\varepsilon^{\mu\nu\rho\sigma}T\indices{^\lambda_{\mu\nu}}T_{\lambda\rho\sigma}+\\
     &+\frac{ (\lambda_2-1)}{2}\varepsilon^{\mu\nu\rho\sigma} T\indices{^\lambda_{\mu\nu}}Q\indices{_{\rho\sigma\lambda}}.
\end{split}
\end{equation}
In addition, we note that it is in general possible to include in \eqref{NY general} a quadratic term in the nonmetricity, 
\begin{equation}
\begin{split}
     &NY_{gen}+\lambda_3\varepsilon^{\mu\nu\rho\sigma} Q\indices{_{\mu\nu}^\lambda}Q\indices{_{\rho\sigma\lambda}}=\\
     =&NY_{gen}+\lambda_3\varepsilon^{\mu\nu\rho\sigma}\Omega\indices{_{\mu\nu}^\lambda}\Omega_{\rho\sigma\lambda},
\end{split}
\label{NY general not top}
\end{equation}
such that by selecting the purely tensor part of nonmetricity is trivially preserved under projective transformations. However, since a term of the form \eqref{NY general not top} does not affect at all the equations for the vector part of the connection, and does not alter the solution for the tensorial part ($\Omega_{\rho\mu\nu}=0$ is still the solution, see Sec.~\ref{section 4}), we can safely omit it from the analysis.
\\ In the following, therefore, we will consider the general form \eqref{NY general}, which by a suitable choice of the parameters $\lambda_{1,2}$ can reproduce all the well known actions usually studied in LQG, as the Holst ($\lambda_1=\lambda_2=0$) or the standard Nieh-Yan \eqref{NY} ($\lambda_1=1,\,\lambda_2=0$) terms.
\section{Generalized Nieh-Yan models}\label{section 4}
\noindent Let us therefore consider for the gravitational sector the action\footnote{We set $\kappa=8\pi G$ and $c=\hbar=1$.}
\begin{equation}
    S_g=\frac{1}{2\kappa}\int d^4x \sqrt{-g}\,F(R,NY_{gen}),
    \label{action general ny}
\end{equation}
where $F$ is a function of the Ricci scalar $R$ and the generalized Nieh-Yan term $NY_{gen}$. Now, if the following holds
\begin{equation}
    \frac{\partial^2 F}{\partial R^2}\frac{\partial^2 F}{\partial NY_{gen}^2}-\frac{\partial^2 F}{\partial R\partial NY_{gen}}\neq 0,
\end{equation}
we can introduce the scalar tensor representation
\begin{equation}
    S_g=\frac{1}{2\kappa}\int d^4x \sqrt{-g}\leri{\phi R+\beta NY_{gen}-W(\phi,\beta)},
    \label{action general scalar tensor ny}
\end{equation}
with $\phi\equiv\frac{\partial F}{\partial R},\,\beta\equiv\frac{\partial F}{\partial NY_{gen}}$ and $W\equiv\phi R(\phi,\beta)+\beta NY_{gen}(\phi,\beta)-F(\phi,\beta)$. 
\\ \noindent Now, it is clear that by considering actions of the form \eqref{action general ny} we are able to generate in a natural way an Immirzi scalar field, denoted by $\beta$, as one of the scalar degrees of freedom emerging from the Jordan frame representation \eqref{action general scalar tensor ny}. This procedure, moreover, offers a viable mechanism able to produce the interaction term $W(\phi,\beta)$. Finally, matter is included in the model by adding to \eqref{action general ny} the action $S_m(g_{\mu\nu},\chi)$, where we denote with $\chi$ generic matter fields which we assume do not couple with the connection and are minimally coupled to the metric.
\\Now, varying \eqref{action general scalar tensor ny} with respect to the connection we get
\begin{align}
    &-\nabla_\lambda\leri{\sqrt{-g}\phi g^{\mu\nu}}+\delta^\nu_\lambda\nabla_\rho\leri{\sqrt{-g}\phi g^{\mu\rho}}+\nonumber\\
    &-\nabla_\rho\leri{\sqrt{-g}\beta \varepsilon\indices{_\lambda^{\rho\mu\nu}}}+\nonumber\\
    &+\sqrt{-g}\phi\leri{g^{\mu\nu}T_{\lambda}-\delta\indices{^\nu_\lambda}T^{\mu}+T\indices{^{\nu\mu}_\lambda}}\nonumber\\
    &+\sqrt{-g}\beta\left[\varepsilon\indices{_\lambda^{\rho\mu\nu}}T_\rho+\frac{1}{2}\varepsilon\indices{_\lambda^{\mu\rho\sigma}}T\indices{^\nu_{\rho\sigma}}-\frac{\lambda_2}{2}\varepsilon\indices{_\lambda^{\nu\rho\sigma}}T\indices{^\mu_{\rho\sigma}}+\right.\nonumber\\
    &\left.+\varepsilon\indices{^{\mu\nu\rho\sigma}}\leri{\leri{\lambda_1-\frac{\lambda_2}{2}}T_{\lambda\rho\sigma}+\lambda_2\,Q_{\rho\sigma\lambda}}\right]=0\label{equation connection Nieh Yan general}
\end{align}
where we used the Palatini identity for the torsional case
\begin{equation}
    \delta R\indices{^\rho_{\mu\sigma\nu}}=\nabla_\sigma\delta\Gamma\indices{^\rho_{\mu\nu}}-\nabla_\nu\delta\Gamma\indices{^\rho_{\mu\sigma}}-T\indices{^\lambda_{\sigma\nu}}\delta\Gamma\indices{^\rho_{\mu\lambda}}
\end{equation}
and the relation
\begin{align}
    &\int d^4 x\; \nabla_\mu\leri{\sqrt{-g} V^\mu}=\int d^4x\; \partial_\mu\leri{\sqrt{-g}V^\mu}\nonumber\\
    &+\int d^4x\; \sqrt{-g}\;T\indices{^\rho_{\mu\rho}} V^\mu.
\end{align}
From \eqref{equation connection Nieh Yan general} one can extract the equations for the four vector components describing torsion and nonmetricity by successive contractions with $\delta\indices{^\lambda_\mu},\,\delta\indices{^\lambda_\nu},\,g_{\mu\nu}$ and $\varepsilon\indices{^\lambda_{\mu\nu\sigma}}$, i.e.:
\begin{equation}
    \begin{cases}
    &(\lambda_1-\lambda_2)S_\mu=0\\
    &Q_\mu-4P_\mu+\frac{\beta}{\phi}(\lambda_1-\lambda_2)S_\mu=0\\
    &Q_\mu-P_\mu+2T_\mu+\frac{\beta}{2\phi}(1-\lambda_1)S_\mu=3\nabla_\mu\ln\phi\\
    &(1-\lambda_2)(Q_\mu-P_\mu)+2(1-\lambda_1)T_\mu-\frac{\phi}{2\beta}S_\mu=3\nabla_\mu\ln\beta.
    \end{cases}
    \label{system of vectors general}
\end{equation}
We see that the system is always characterized by Weyl geometry configurations, namely by $Q_\mu=4P_\mu$, which allows to recast nonmetricity in the simpler form $Q_{\rho\mu\nu}=P_\rho g_{\mu\nu}+\Omega_{\rho\mu\nu}$. The system \eqref{system of vectors general} can be put therefore in the form
\begin{equation}
    \begin{cases}
    &(\lambda_1-\lambda_2)S_\mu=0\\
    &Q_\mu=4P_\mu\\
    &3P_\mu+2T_\mu+\frac{\beta}{2\phi}(1-\lambda_1)S_\mu=3\nabla_\mu\ln\phi\\
    &3(1-\lambda_2)P_\mu+2(1-\lambda_1)T_\mu-\frac{\phi}{2\beta}S_\mu=3\nabla_\mu\ln\beta.
    \end{cases}
    \label{system of vectors general 2}
\end{equation}
Eventually, we can extract from \eqref{equation connection Nieh Yan general} the last equation for the tensor part encoded in $q_{\rho\mu\nu}$ and $\Omega_{\rho\mu\nu}$, which after quite lengthy calculations results in
\begin{equation}
\begin{split}
     &\frac{\phi}{\beta}\leri{\Omega_{\lambda\mu\nu}-q_{\nu\mu\lambda}}=\\
     &=\frac{1}{2}\varepsilon\indices{^{\rho\sigma}_{\lambda\mu}}q_{\nu\rho\sigma}-\frac{\lambda_2}{2}\varepsilon\indices{^{\rho\sigma}_{\lambda\nu}}q_{\mu\rho\sigma}+\\
     &+\varepsilon\indices{^{\rho\sigma}_{\mu\nu}}\leri{\leri{\lambda_1-\frac{\lambda_2}{2}}q_{\lambda\rho\sigma}+(\lambda_2-1)\Omega_{\rho\sigma\lambda}}
\end{split}\label{equation tensor modes}
\end{equation}
where we used repeatedly \eqref{system of vectors general 2}. Then, by taking the symmetric part of \eqref{equation tensor modes} in the indices $\mu,\nu$ we can express the nonmetricity 3-rank part in terms of the torsional analog, i.e.
\begin{equation}
    \phi\,\Omega_{\lambda\mu\nu}=\phi\, q_{(\nu\mu)\lambda}+\frac{\beta (1-\lambda_2)}{2}\varepsilon\indices{^{\rho\sigma}_{\lambda(\mu}}q_{\nu)\rho\sigma},
\end{equation}
which inserted back in \eqref{equation tensor modes} leads to the trivial solution for the tensor modes $\Omega_{\lambda\mu\nu}=q_{\nu\mu\lambda}=0$. In the following, therefore, we can only focus on the purely vector modes \eqref{system of vectors general 2}, and we see that the structure of the solution depends crucially on the relation between the parameter $\lambda_1$ and $\lambda_2$. When projective invariance is explicitly broken, as it occurs for $\lambda_1\neq\lambda_2$, we are compelled to set $S_\mu=0$ and the general solution is displayed by
\begin{equation}
    \begin{cases}
    &S_\mu=0\\
    &Q_\mu=4P_\mu\\
    &P_\mu=\frac{1}{\lambda_1-\lambda_2}\bar{\nabla}_\mu\ln\beta+\frac{\lambda_1-1}{\lambda_1-\lambda_2}\bar{\nabla}_\mu\ln\phi\\
    &T_\mu=-\frac{3}{2}\frac{1}{\lambda_1-\lambda_2}\bar{\nabla}_\mu\ln\beta-\frac{3}{2}\frac{\lambda_2-1}{\lambda_1-\lambda_2}\bar{\nabla}_\mu\ln\phi\\
    &q_{\rho\mu\nu}=\Omega_{\rho\mu\nu}=0.
    \end{cases}
    \label{solution vector projective breaking}
\end{equation}
If $\lambda_1=\lambda_2\equiv\lambda$, instead, we have at our disposal the projective invariance for getting rid of one vector degree of freedom, which can be made vanishing by properly setting the vector $\xi_\mu$. We can decide, for instance, to set $\xi_\mu=-\frac{1}{2}P_\mu$, in order to deal in \eqref{system of vectors general 2} only with torsion\footnote{For the sake of clarity we omit the tilde notation for transformed quantities.}. We obtain then:
\begin{equation}
    \begin{cases}
    &Q_\mu=4P_\mu=0\\
    &S_\mu=\frac{6\beta(1-\lambda)}{\beta^2(1-\lambda)^2+\phi^2}\bar{\nabla}_\mu\phi-\frac{6\phi}{\beta^2(1-\lambda)^2+\phi^2}\bar{\nabla}_\mu\beta\\
    &T_\mu=\frac{3}{2}\frac{\phi}{\beta^2(1-\lambda)^2+\phi^2}\bar{\nabla}_\mu\phi+\frac{3}{2}\frac{\beta(1-\lambda)}{\beta^2(1-\lambda)^2+\phi^2}\bar{\nabla}_\mu\beta\\
    &q_{\rho\mu\nu}=\Omega_{\rho\mu\nu}=0.
    \end{cases}
    \label{solution vector projective preserving}
\end{equation}
We remark that while in \eqref{solution vector projective breaking} the affine structure is strictly fixed, leading to the presence of torsion and nonmetricity, in \eqref{solution vector projective preserving}, as a matter of fact, we could have chosen $\xi_\mu=\frac{T_\mu}{3}$ and retained the nonmetricity vector $P_\mu$ instead of the torsion trace. Such a flexibility in the specific representation of the theory, however, does not reflect in a dynamical vagueness, and the proper degrees of freedom can be unambiguously identified. Let us re-express \eqref{NY general}, indeed, in terms of its vector components, i.e.
\begin{equation}\label{NYgen components}
    NY_{gen}=-\frac{1}{2}\bar{\nabla}\cdot S-\frac{(1-\lambda_1)}{3}S \cdot T-\frac{(1-\lambda_2)}{2}S\cdot P.
\end{equation}
Then, looking at \eqref{solution vector projective breaking}, it is clear that the solution $S_\mu=0$ remarkably implies that the generalized Nieh-Yan term \eqref{NY general} is identically vanishing on half-shell. In other words, the theory can rearrange its affine structure in such a way that terms violating projective invariance be harmless along the dynamics. This can be further appreciated by looking at the effective scalar tensor action stemming from \eqref{action general scalar tensor ny}, when \eqref{solution vector projective breaking} are plugged in it\footnote{We could have also varied \eqref{action general scalar tensor ny} with respect to the other degrees of freedom, and then inserted the solutions for connections. Since in this case we would obtain the same equations of motion for the metric and the scalar fields, for the sake of clarity we chose to deal directly with \eqref{NY scalar tensor}.}. Explicit calculations lead to
\begin{equation}
    S=\frac{1}{2\kappa}\int d^4x \sqrt{-g}\leri{\phi \bar{R}+\frac{3}{2\phi}\bar{\nabla}_\mu\phi\bar{\nabla}^\mu\phi-W(\phi,\beta)},
    \label{NY scalar tensor}
\end{equation}
which resembles the form of a Palatini $f(R)$ theory, with a potential depending on two scalar fields. In particular, we see that in this case the equation for the Immirzi field is simply given by
\begin{equation}
    \frac{\partial W(\phi,\beta)}{\partial \beta}=0,
    \label{NY beta}
\end{equation}
which actually fixes the form of the Immirzi field in terms of the scalaron $\phi$, i.e. $\beta=\beta(\phi)$. Then, it can be easily verified that the variation of \eqref{NY scalar tensor} with respect to $\phi$, combined with the trace of the equation for the metric field, results in the canonical structural equation featuring Palatini $f(R)$ theories \cite{Olmo2011}, i.e.
\begin{equation}\label{structural NY}
\left[2W(\phi,\beta)-\phi\frac{\partial W(\phi,\beta)}{\partial\phi}\right]_{\beta=\beta(\phi)}=\kappa T,
\end{equation}
\newline
which shows that the dynamics of the scalaron $\phi$ is frozen as well, and completely determined by the trace of the stress energy tensor of matter. Conditions \eqref{NY beta} and \eqref{structural NY} then establish that the scalar fields $\phi,\;\beta$ are not truly propagating degrees of freedom, and reduce to constants in vacuum, where the theory is stable and the breaking of projective invariance does not lead to ghost instabilities as in \cite{BeltranJimenez2019}.
\\When we set $\lambda_1=\lambda_2=\lambda$, instead, with a bit of effort the effective action stemming from \eqref{solution vector projective preserving} can be rearranged in the form
\begin{widetext}
\begin{equation}
    S=\frac{1}{2\kappa}\int d^4 x \sqrt{-g} \leri{\phi \bar{R}+\frac{3\phi}{2(\beta^2(1-\lambda)^2+\phi^2)}\leri{\bar{\nabla}_\mu\phi\bar{\nabla}^\mu\phi-\bar{\nabla}_\mu\beta\bar{\nabla}^\mu\beta+\frac{2\beta(1-\lambda)}{\phi}\bar{\nabla}_\mu\phi\bar{\nabla}^\mu\beta}-W(\phi,\beta)},
\end{equation}
where the mixing term $\bar{\nabla}_\mu\phi\bar{\nabla}^\mu\beta$ can be always reabsorbed by the transformation\footnote{We see that in the special case of $\lambda=1$, when also topologicity is restored, no redefinition for the Immirzi field is required and his kinetic term simply boils down to $-\frac{3}{2\phi}(\nabla\beta)^2$.} $\psi\equiv\beta\phi^{\lambda-1}$, which puts the action in the diagonal form:
\begin{equation}
    S=\frac{1}{2\kappa}\int d^4 x \sqrt{-g} \leri{\phi \bar{R}+\frac{3}{2\phi}\bar{\nabla}_\mu\phi\bar{\nabla}^\mu\phi-\frac{3\phi}{2}\frac{1}{\phi^{2\lambda}+(1-\lambda)^2\psi^2}\bar{\nabla}_\mu\psi\bar{\nabla}^\mu\psi-V(\phi,\psi)},
    \label{effective action projective preserving}
\end{equation}
\end{widetext}
where we redefined $V(\phi,\psi)=W(\phi,\psi \phi^{1-\lambda})$. It is clear, therefore, that we expect in general the Immirzi field to be a well-behaved dynamical degree of freedom, as it can be appreciated in the Einstein frame defined by the conformal rescaling $\tilde{g}_{\mu\nu}=\phi\,g_{\mu\nu}$ (see \cite{Fujii2003,Olmo:2005hc} for details). In the Einstein frame the nonminimal coupling of $\phi$ with the Ricci scalar is removed along with its kinetic term, and we can just look at the kinetic term for the Immirzi field which takes the form
\begin{equation}
    -\frac{3}{2}\frac{\tilde{g}^{\mu\nu}\nabla_\mu\psi\nabla_\nu\psi}{\phi^{2\lambda}+(1-\lambda)^2\psi^2}.
    \label{noghost}
\end{equation}
Now, since the inequality $\phi^{2\lambda}+(1-\lambda)^2\psi^2>0$ holds irrespective of the values of $\phi,\,\psi$ and $\lambda$, \eqref{noghost} has always the correct sign and no ghost instability arises.

\subsection{Comparison with Riemann-Cartan solutions}
We saw that when projective invariance is conserved, as a matter of fact nonmetricity can be entirely neglected by properly fixing the vector $\xi_\mu$. This seems to suggest, similarly to what outlined in \cite{Iosifidis:2018zjj,Afonso:2017bxr}, a duality between torsion and nonmetricity when $f(R)$-like extensions of general relativity are considered. In this respect, therefore, it is interesting to analyze the structure of the solutions for vanishing nonmetricity, when this property is not the result of a projective transformation but a preliminary condition we impose on the metric-affine structure. 
\\ When we simply disregard nonmetricity contributions in \eqref{system of vectors general 2}, we are just selecting a particular subset of solutions for the models violating projective invariance, i.e. 
\begin{equation}
    \begin{cases}
    &S_\mu=0\\
    &T_\mu=\frac{3}{2\phi}\bar{\nabla}_\mu\phi\\
    &\bar{\nabla}_\mu\beta=0,
    \end{cases}
    \label{nonmetr by hand}
\end{equation}
which, coherently, leads again to the effective action \eqref{NY scalar tensor}. Now, however, we are compelled to select a constant Immirzi parameter, and by virtue of \eqref{NY beta} the last one of \eqref{nonmetr by hand} simply implies
\begin{equation}
    \partial_\mu\beta=\frac{\partial\beta}{\partial\phi}\partial_\mu \phi(T)=0,
\end{equation}
which for a generic $T\neq 0$ is satisfied if $\partial\beta/\partial\phi=0$, that is to say whenever the potential $W(\phi,\beta)$ does not depend on $\beta$. This requirement eliminates the Immirzi parameter from \eqref{NY scalar tensor} and fully restores the equivalence of the form of the action with the Palatini $f(R)$ gravity.
\\The Riemann-Cartan structure of \cite{Calcagni2009,Mercuri2006,Mercuri2008,Mercuri2009,Mercuri2009a,Bombacigno2016,Bombacigno2018,Bombacigno2019} can be instead properly replicated by implementing in \eqref{action general ny} the condition of vanishing nonmetricity with a Lagrange multiplier, i.e.
\begin{equation}
     S_g^{RC}=\frac{1}{2\kappa}\int d^4x \sqrt{-g}\,\left[F(R,NY_{gen})+l^{\rho\mu\nu}Q_{\rho\mu\nu}\right],
\end{equation}
where $l^{\rho\mu\nu}=l^{\rho\nu\mu}$. In so doing, indeed, we are not forcing $S_\mu=0$, because of the appearance of traces of $l^{\rho\mu\nu}$ in \eqref{system of vectors general 2}, which for $Q_{\rho\mu\nu}=0$ takes the form
\begin{equation}
    \begin{cases}
    &q_\mu=\frac{\beta}{2}(\lambda_1-\lambda_2) S_\mu\\
    &p_\mu=-\frac{1}{2}q_\mu\\
    &S_\mu=\frac{6\beta(1-\lambda_1)}{\beta^2(1-\lambda_1)^2+\phi^2}\bar{\nabla}_\mu\phi-\frac{6\phi}{\beta^2(1-\lambda_1)^2+\phi^2}\bar{\nabla}_\mu\beta\\
    &T_\mu=\frac{3}{2}\frac{\phi}{\beta^2(1-\lambda_1)^2+\phi^2}\bar{\nabla}_\mu\phi+\frac{3}{2}\frac{\beta(1-\lambda_1)}{\beta^2(1-\lambda_1)^2+\phi^2}\bar{\nabla}_\mu\beta,
    \end{cases}
\end{equation}
where the traces $q_\mu\equiv l\indices{_\mu^\rho_\rho}$ and $p_\mu\equiv l\indices{^\rho_{\mu\rho}}$ are completely solved in terms of the axial vector $S_\mu$. Then, results of \cite{Calcagni2009,Mercuri2006,Mercuri2008,Mercuri2009,Mercuri2009a,Bombacigno2016,Bombacigno2018,Bombacigno2019} are simply obtained\footnote{Obviously, the parameter $\lambda_2$ does not appear at all in the expressions for the vectors.} by setting $\lambda_1=1$, and \eqref{effective action projective preserving} reproduced.

\section{Dynamical Immirzi models}\label{secIV}
Here we focus on models described by \eqref{effective action projective preserving}, which we saw to be endowed with a dynamical Immirzi field. Then, let us evaluate the equation of motion for the metric
\begin{equation}
\begin{split}
    &\bar{G}_{\mu\nu}=\frac{\kappa}{\phi}T_{\mu\nu}+\frac{1}{\phi}\leri{\bar{\nabla}_\mu\bar{\nabla}_\nu-g_{\mu\nu}\bar{\Box}}\phi+\\&-\frac{3}{2\phi^2}\bar{\nabla}_\mu\phi\bar{\nabla}_\nu\phi+\frac{3}{2}\frac{\bar{\nabla}_\mu\psi\bar{\nabla}_\nu\psi}{\phi^{2\lambda}+(1-\lambda)^2\psi^2}\\
    &+\frac{1}{2}g_{\mu\nu}\leri{\frac{3(\bar{\nabla}\phi)^2}{2\phi^2}-\frac{3}{2}\frac{(\bar{\nabla}\psi)^2}{\phi^{2\lambda}+(1-\lambda)^2\psi^2}-\frac{V(\phi,\psi)}{\phi}},
    \label{equation DI metric}
\end{split}
\end{equation}
and the scalar fields, i.e.
\begin{align}
    &2V(\phi,\psi)-\phi\frac{\partial V(\phi,\psi)}{\partial\phi}\nonumber\\
    &+\frac{3\lambda \phi^{2\lambda+1}}{\leri{\phi^{2\lambda}+(1-\lambda)^2\psi^2}^2}(\bar{\nabla}\psi)^2=\kappa T,
    \label{structural DI}
    \\
    &\bar{\Box}\psi-\frac{(1-\lambda)^2\psi}{\phi^{2\lambda}+(1-\lambda)^2\psi^2}(\bar{\nabla}\psi)^2+\nonumber\\
    &\leri{1-\frac{2\lambda\phi^{2\lambda}}{\phi^{2\lambda}+(1-\lambda)^2\psi^2}}\bar{\nabla}_\mu\ln\phi\bar{\nabla}^\mu\psi=\frac{\partial V(\phi,\psi)}{3\partial\psi},
    \label{equation DI immirzi}
\end{align}
where \eqref{structural DI} is obtained in analogy with \eqref{structural NY}. From the first equation we see that the scalaron $\phi$ can be algebraically solved in terms of the Immirzi field and its kinetic term $X\equiv (\bar{\nabla}\psi)^2$, i.e.
\begin{equation}
    \phi=\phi(\psi,X,T),
    \label{solution phi DI}
\end{equation}
so that we are left with an only propagating degree of freedom, the Immirzi field. Moreover, equation \eqref{solution phi DI} suggests an intriguing analogy with the so called Degenerate Higher-Order Scalar-Tensor (DHOST) theories \cite{Langlois2016,BenAchour2019}, where higher order derivatives of the scalar field in the action do not actually lead to dynamical instabilities, by virtue of some degeneracy conditions on the kinetic matrix. An important subclass of DHOST theories is that one in agreement with the absence of graviton decay and the experimental constraint on the speed of gravitational
waves \cite{Langlois2018}, which are described by the action
\begin{align}\label{DHOST exp comp}
S_{DHOST}&=\frac{1}{2\kappa}\int d^4x \sqrt{-g} \left[ F_0 + F_1 \bar{\Box}\varphi + F_2 \bar{R}  \right.\nonumber\\
&\left.+\frac{6F_{2X}^2}{F_2}\varphi^{\mu}\varphi^{\nu}\varphi_{\mu\lambda}\varphi\indices{^{\lambda}_{\nu}} \right],
\end{align}
where $\varphi_\mu \equiv \nabla_\mu \varphi,\,\varphi_{\mu\nu} \equiv \nabla_\mu \nabla_\nu\varphi$, and $F_0$, $F_1$, $F_2$ are functions of the kinetic term $X\equiv \varphi^{\mu}\varphi_{\mu}$.
In this regard, then, consider for \eqref{action general ny} the $F(R)+NY^*$ model in vacuum ($T=0$), identified by $\lambda=1$ and the condition $\partial V/\partial \psi=0$ (or $V_2(\psi)=0$, see below). In this case equation \eqref{solution phi DI} simply reads $\phi=\phi(X)$. Direct substitution of the latter into \eqref{effective action projective preserving} yields the equivalence at the Lagrangian level with \eqref{DHOST exp comp}, upon identification of the DHOST scalar field $\varphi$ with the Immirzi field and considering the following functional choices:
\begin{subequations}
\begin{align}
F_0&=-V(\phi(X))-\frac{3X}{2\phi(X)},\\
F_1&=0,\\
F_2&=\phi(X).
\end{align}\label{DHOST potentials}
\end{subequations}
In particular, the requirement that the field equations stemming from \eqref{DHOST exp comp} and \eqref{action general ny} be equivalent, leads to the additional condition
\begin{equation}\label{condition equivalence DHOST}
X - \frac{\phi(X)}{\phi_X(X)} \neq 0,
\end{equation}
which rules out the linear case $\phi(X)\propto X$. We note, in addition, that for the subclass \eqref{DHOST exp comp}, the degeneracy condition preventing the arising of Ostrogradsky instabilities simply reads $F_2(X)\neq 0$, which is consistent with the requirement $\phi\neq 0$.
\\Moreover, it is interesting to note that the dependence of $\phi$ on the trace $T$ of the stress energy tensor, which holds in general for projective invariant models, introduces a dependence of the affine structure on the matter, even if we assumed at the beginning a vanishing hypermomentum, i.e.
\begin{equation}
    \Delta\indices{_\lambda^{\mu\nu}}\equiv -\frac{2}{\sqrt{-g}}\frac{\delta S_M}{\delta\Gamma\indices{^\lambda_{\mu\nu}}}=0.
\end{equation}
This, possibly, suggests a mechanism for circumventing the inconsistencies which usually arise when one tries to implement symmetries, like the projective invariance, in the presence of matter fields which couple to the  connection \cite{Iosifidis:2019fsh}.
\\Finally, we see that at the first order in perturbation \eqref{solution phi DI} implies in vacuum $\delta\phi\sim\delta\psi$, and the inspection of \eqref{equation DI metric} suggests that in this case the Immirzi field could actually mimic the scalar polarization of gravitational waves in metric $f(R)$ gravity (see \cite{Moretti2019,Liang2017} for details). On the other hand, observations on gravitational waves propagation \cite{Abbott:2017vtc} and Solar System dynamics \cite{Berry2011,Berry2012} put severe constraints on the mass of additional scalar degrees of freedom, which for many purposes can be satisfactorily considered massless. Since this amounts to disregarding the potential term in \eqref{equation DI immirzi}, it makes sense to seek for a subclass of functions $F(R,NY_{gen})$ able to generate separable potentials $V(\phi,\psi)=V_1(\phi)+V_2(\psi)$, under the assumption that $V_2(\psi)$ can be safely neglected. At first sight, a vanishing Immirzi potential may conflict with the requirement of reproducing standard LQG predictions, as it can occur whenever the Immirzi field collapses on a minimum configuration. Below, however, we demonstrate that this is not actually mandatory, since the dynamics of the Immirzi scalar can be adequately frozen by cosmological evolution as well, featured by classical big bounce scenarios.

\section{Big bounce in Bianchi I cosmology}\label{section 6}

\noindent Let us set $\lambda=1$, corresponding to the projectively invariant Nieh-Yan model, and fix the form of the potential as $V(\phi,\psi)=V(\phi)$. In this case, the equations of motion can be rearranged as:
\begin{equation}
    \begin{split}
        \bar{G}_{\mu\nu}&=\frac{\kappa}{\phi}T_{\mu\nu}+\frac{1}{\phi}\leri{\bar{\nabla}_\mu\bar{\nabla}_\nu-g_{\mu\nu}\bar{\Box}}\phi+\\
        &-\frac{3}{2\phi^2}\bar{\nabla}_\mu\phi\bar{\nabla}_\nu\phi+\frac{3}{2\phi^2}\bar{\nabla}_\mu\psi\bar{\nabla}_\nu\psi+\\
        &+\frac{1}{2}g_{\mu\nu}\leri{\frac{3}{2\phi^2}(\bar{\nabla}\phi)^2-\frac{3}{2\phi^2}(\bar{\nabla}\psi)^2-\frac{V(\phi)}{\phi}},
    \label{equation DI metric ny}
    \end{split}
\end{equation}
and
\begin{align}
    &2V(\phi)-\phi\frac{d V(\phi)}{d\phi}=\kappa T-\frac{3(\bar{\nabla}\psi)^2}{\phi},
    \label{structural DI ny}
    \\
    &\bar{\Box}\psi-\bar{\nabla}_\mu\ln\phi\bar{\nabla}^\mu\psi=0.
    \label{equation DI immirzi ny}
\end{align}
Now, we consider the metric for a Bianchi I flat spacetime, i.e.
\begin{equation}
    ds^2=-dt^2+a(t)^2dx^2+b(t)^2dy^2+c(t)^2dz^2,
    \label{bianchi metric}
\end{equation}
 which represents the simplest example of homogeneous spacetime endowed with anisotropies, encoded in the three different scale factors $a(t),b(t),c(t)$. We assume, moreover, that matter is described by a perfect fluid, whose stress energy tensor in the comoving frame is given by
\begin{equation}
    T_{\mu\nu}=\text{diag}(\rho, a^2p,b^2p,c^2p),
    \label{perfect fluid tensor}
\end{equation}
where $\rho$ is the energy density and $p$ the pressure. Then, it is easy to check that it is covariantly conserved, i.e. $\bar{\nabla}_\mu T^{\mu\nu}=0$, leading to the continuity equation\footnote{We denote with a dot derivatives with respect to the coordinate time $t$.}
\begin{equation}
    \dot{\rho}+\leri{\frac{\dot{a}}{a}+\frac{\dot{b}}{b}+\frac{\dot{c}}{c}}(\rho+p)=0,
\end{equation}
which for a equation of state of the form $p=w\rho$ results in
\begin{equation}\label{energy density}
    \rho(t)=\frac{\mu^2}{(abc)^{w+1}},
\end{equation}
where $\mu^2$ is a constant. Lastly, in accordance to what we discussed in Sec.~\ref{secIV}, we take for the function $F(R,NY_{gen})$ an effective form $F(R,NY_{gen})\simeq R+\alpha R^2+NY_{gen}$, which amounts to considering the Starobinsky quadratic potential \cite{STAROBINSKY198099}:
\begin{equation}
    V(\phi)=\frac{1}{\alpha}\leri{\frac{\phi-1}{2}}^2.
    \label{potential quadratic}
\end{equation}
Thus, we observe that with the metric \eqref{bianchi metric} the equation for the Immirzi field \eqref{equation DI immirzi ny} can be solved analytically for $\dot{\psi}$ taking the form
\begin{equation}
    \dot{\psi}=\frac{k_0\phi}{abc},
    \label{solution immirzi DI hy}
\end{equation}
which plugged into \eqref{structural DI ny}, and by taking into account \eqref{perfect fluid tensor} and \eqref{potential quadratic}, allows us to express the field $\phi$ in terms of scale factors as
\begin{equation}
    \phi=\frac{v^2 f(v)}{6\alpha k_0^2+v^2},
    \label{solution phi DI ny}
\end{equation}
where we introduced the volume-like variable $v\equiv abc$ and the function $f(v)\equiv 1-2\alpha\kappa(3w-1)\rho(v)$. It follows that the only non vanishing elements of \eqref{equation DI metric ny} are the $tt,xx,yy$ and $zz$ components, which take the form, respectively
\begin{align}
    \frac{\dot{a}\dot{b}}{ab}+\frac{\dot{a}\dot{c}}{ac}+\frac{\dot{b}\dot{c}}{bc}&=\frac{\kappa\rho}{\phi}+\frac{3k_0^2}{4v^2}-\leri{\frac{\dot{a}}{a}+\frac{\dot{b}}{b}+\frac{\dot{c}}{c}}\frac{\dot{\phi}}{\phi}\nonumber\\
    &-\frac{3\dot{\phi}^2}{4\phi^2}+\frac{V(\phi)}{2\phi},
    \label{tt component ny}\\
    \frac{\ddot{b}}{b}+\frac{\ddot{c}}{c}+\frac{\dot{b}\dot{c}}{bc}&=-\leri{\frac{\dot{b}}{b}+\frac{\dot{c}}{c}}\frac{\dot{\phi}}{\phi}+\Phi,
    \label{xx component ny}\\
    \frac{\ddot{a}}{a}+\frac{\ddot{c}}{c}+\frac{\dot{a}\dot{c}}{ac}&=-\leri{\frac{\dot{a}}{a}+\frac{\dot{c}}{c}}\frac{\dot{\phi}}{\phi}+\Phi,
    \label{yy component ny}\\
    \frac{\ddot{a}}{a}+\frac{\ddot{b}}{b}+\frac{\dot{a}\dot{b}}{ab}&=-\leri{\frac{\dot{a}}{a}+\frac{\dot{b}}{b}}\frac{\dot{\phi}}{\phi}+\Phi.
    \label{zz component ny}
\end{align}
where
\begin{equation}
   \Phi \equiv -\frac{\kappa p}{\phi}-\frac{3k_0^2}{4v^2}-\frac{\ddot{\phi}}{\phi}+\frac{3\dot{\phi}^2}{4\phi^2}+\frac{V(\phi)}{2\phi}.
\end{equation}
The purpose of our analysis is now to rearrange \eqref{tt component ny}, which in the limit $a=b=c$ reproduces the Friedmann equation of the scale factor for the FRW Universe, in such a way that its l.h.s. is manifestly positive and the r.h.s. displays  a rational function in $v$. This allows us to qualitatively determine the behaviour of $v$ by means of algebraic techniques, since the existence of singularities, turning points, or big-bounce scenarios can be related with the zeros and the poles of the function in the r.h.s.. In order to see that, we start by noting that combining \eqref{yy component ny} with \eqref{zz component ny}, we get
\begin{equation}
    \frac{\ddot{b}}{b}-\frac{\ddot{c}}{c}+\leri{\frac{\dot{b}}{b}-\frac{\dot{c}}{c}}\frac{\dot{a}}{a}=-\leri{\frac{\dot{b}}{b}-\frac{\dot{c}}{c}}\frac{\dot{\phi}}{\phi},
\end{equation}
which can be solved for $a$ as
\begin{equation}
    a=\frac{k_1}{\phi(\dot{b}c-b\dot{c})},
    \label{a solved}
\end{equation}
with $k_1$ an integration constant. Analogously, similar relations can be derived for the other scale factors, which take the form
\begin{align}
   &b=\frac{k_2}{\phi(\dot{a}c-a\dot{c})},
   \label{b solved}\\
   &c=\frac{k_3}{\phi(\dot{a}b-a\dot{b})},
   \label{c solved}
\end{align}
resulting in the constraint $k_1-k_2+k_3=0$. We introduce thus the Hubble-like functions
\begin{equation}
    H_A=\frac{\dot{a}}{a},\;H_B=\frac{\dot{b}}{b},\;H_C=\frac{\dot{c}}{c},
\end{equation}
in terms of which we can rearrange \eqref{a solved}-\eqref{c solved} as
\begin{equation}
    H_B-H_C=\frac{k_1}{\phi v},\;H_A-H_C=\frac{k_2}{\phi v},\;H_A-H_B=\frac{k_2}{\phi v},
\end{equation}
so that they can be combined to give
\begin{equation}
    H_AH_B+H_AH_C+H_BH_C=H_A^2+H_B^2+H_C^2-\frac{3\mu_A^2}{\phi^2 v^2},
    \label{sum hubble rate}
\end{equation}
where we defined the anisotropy density parameter $\mu_A^2\equiv \frac{k_1^2+k_2^2+k_3^2}{6\kappa}$ for future convenience.
Next, we convert the time derivative of the scalaron $\phi$ appearing in \eqref{tt component ny} in a function of $v$, i.e.
\begin{align}
    \frac{\dot\phi}{\phi}&=\frac{\dot{v}}{\phi}\frac{d\phi}{dv},
    \label{chain phi v}
\end{align}
and we note that
\begin{equation}
   \frac{\dot{v}}{v}=H_A+H_B+H_C,
\end{equation}
so that we can write
\begin{equation}
    H_A^2+H_B^2+H_C^2=\leri{\frac{\dot{v}}{v}}^2-2(H_AH_B+H_AH_C+H_BH_C).
    \label{hubble rate volume}
\end{equation}
Then, taking into account \eqref{sum hubble rate}, \eqref{chain phi v} and \eqref{hubble rate volume}, we can finally rearrange \eqref{tt component ny} in the simple form
\begin{equation}
    H^2\equiv\leri{\frac{\dot{v}}{3v}}^2=\frac{\frac{\kappa}{3}\leri{\frac{\mu_I^2}{v^2}+\frac{\rho}{\phi }+\frac{\mu_{AN}^2}{\phi^2v^2}}+\frac{V(\phi)}{6\phi}}{\leri{1+\frac{3v}{2}\frac{d}{dv}\ln\phi}^2},
    \label{hubble function ny}
\end{equation}
where we introduced the energy density parameter for the Immirzi field $\mu^2_I\equiv\frac{3k_0^2}{4\kappa}$, and we finally observe that the r.h.s. is a rational function of $v$.

\onecolumngrid

\begin{figure}[t!]
\centering
\begin{subfigure}[t]{.5\textwidth}
  \centering
  \caption{\label{fig: PlotV} Universe volume normalised to $v_B$.}
  \includegraphics[height=2in]{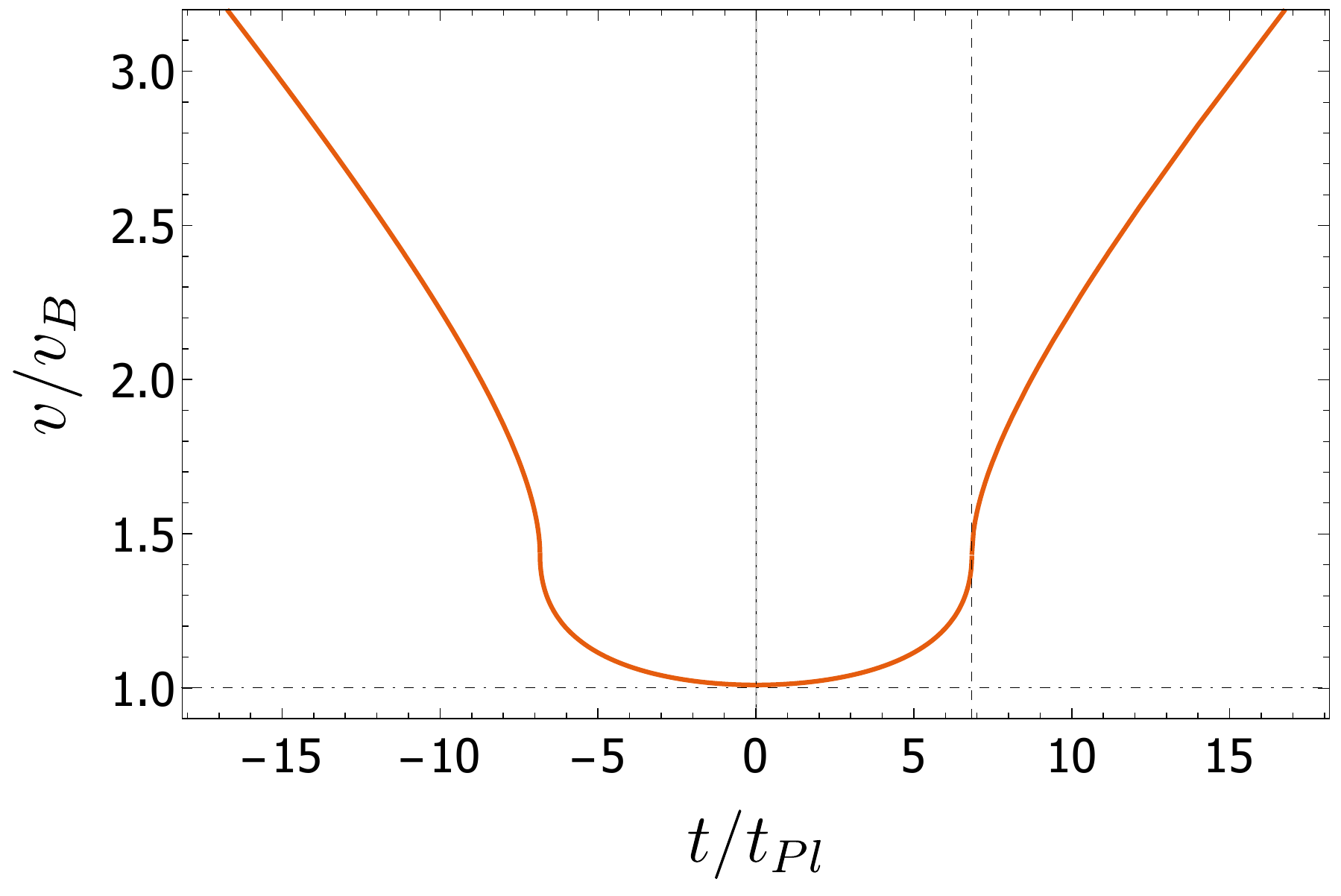}
\end{subfigure}%
 ~
\begin{subfigure}[t]{.5\textwidth}
  \centering
  \caption{\label{fig: PlotScalefactors}Scale factors.}
  \includegraphics[height=2in]{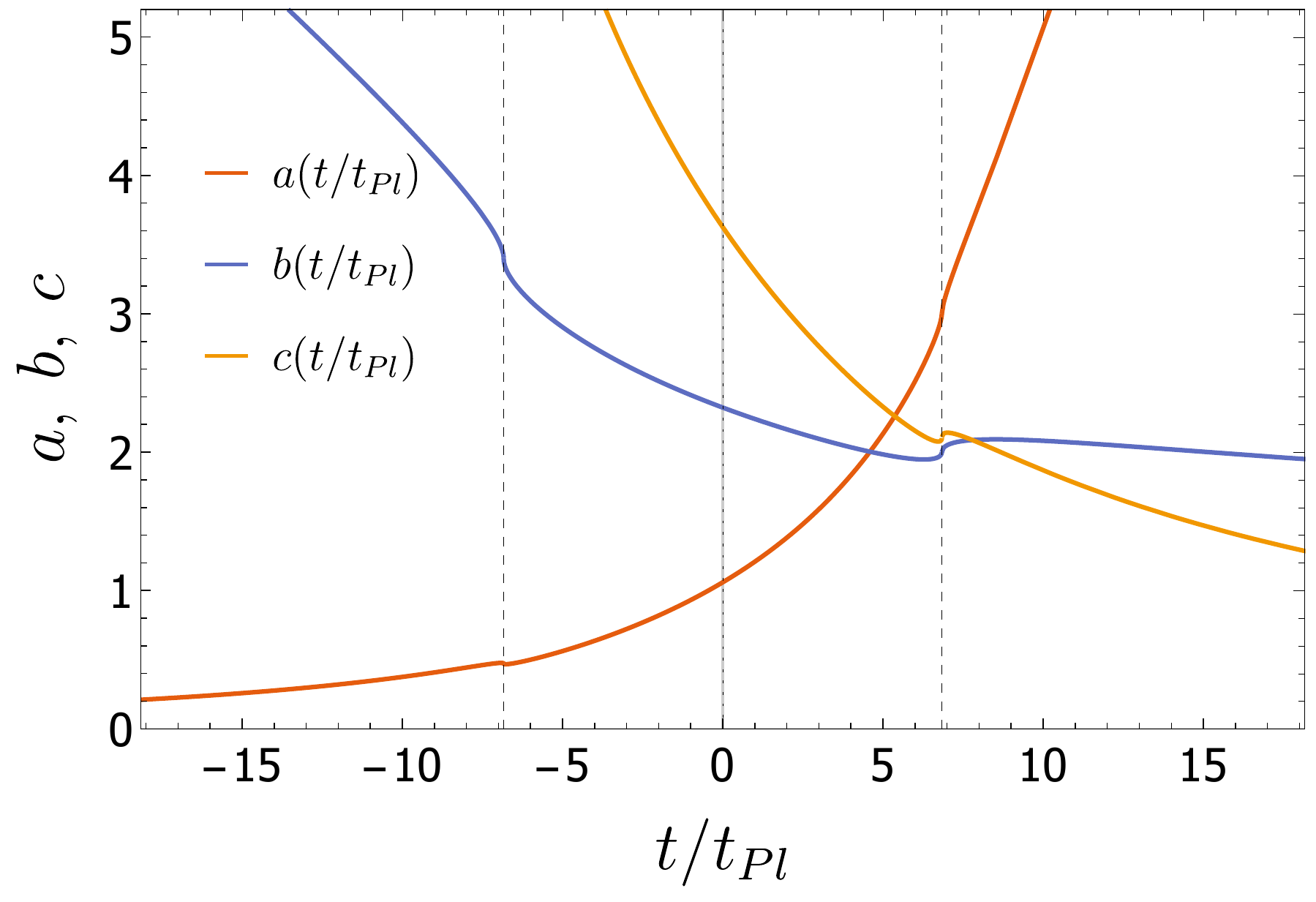}
\end{subfigure}
\newline
\begin{subfigure}[t]{.5\textwidth}
  \centering
  \caption{\label{fig: PhiPsi} Scalaron $\phi$ and Immirzi field derivative $\dot{\psi}$.}
  \includegraphics[height=2in]{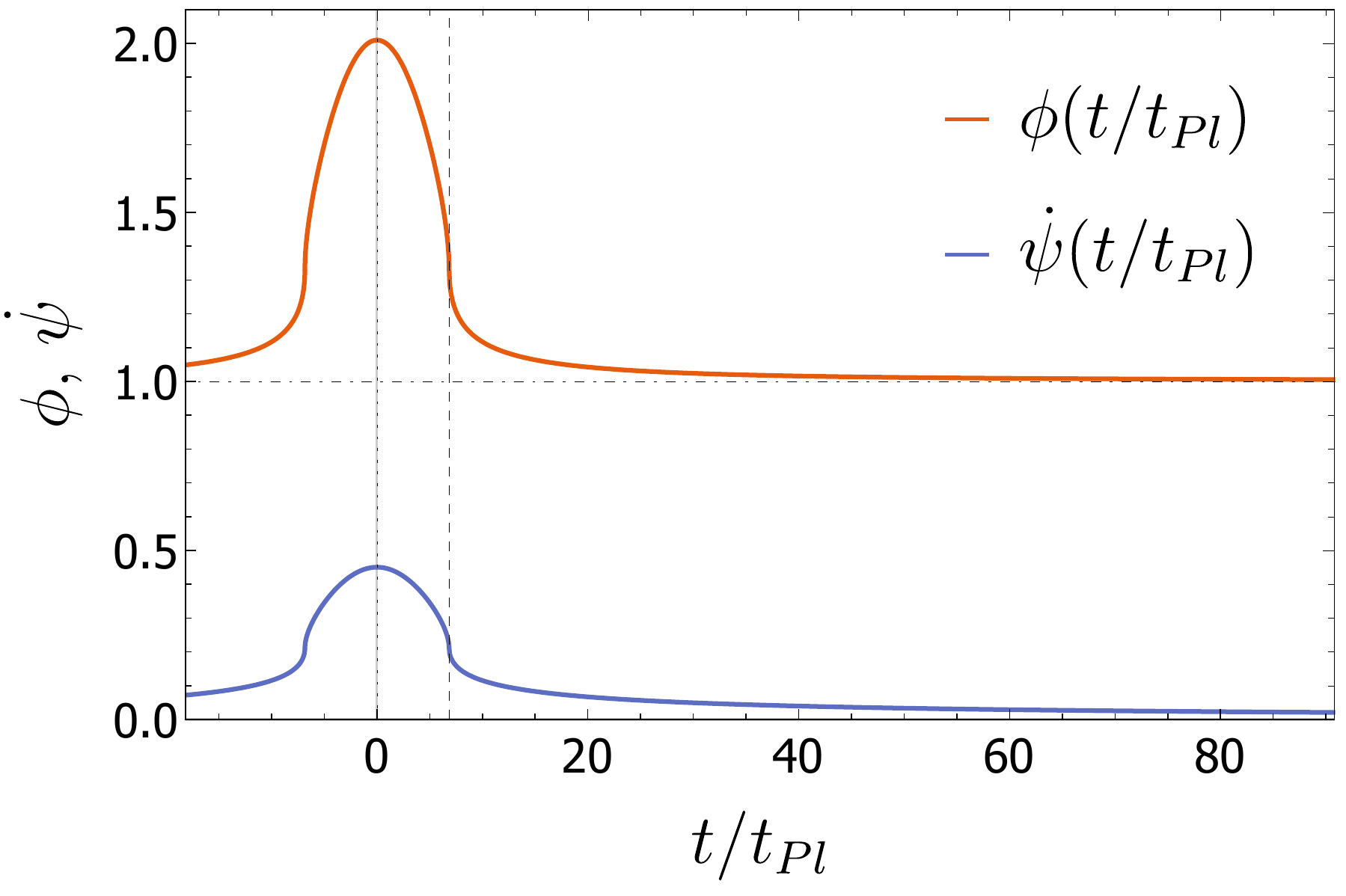}
\end{subfigure}%
 ~
\begin{subfigure}[t]{.5\textwidth}
  \centering
    \caption{\label{fig: BounceAlpha} Value of the bounce volume $v_B$ as a function of $\alpha$.}
  \includegraphics[height=2in]{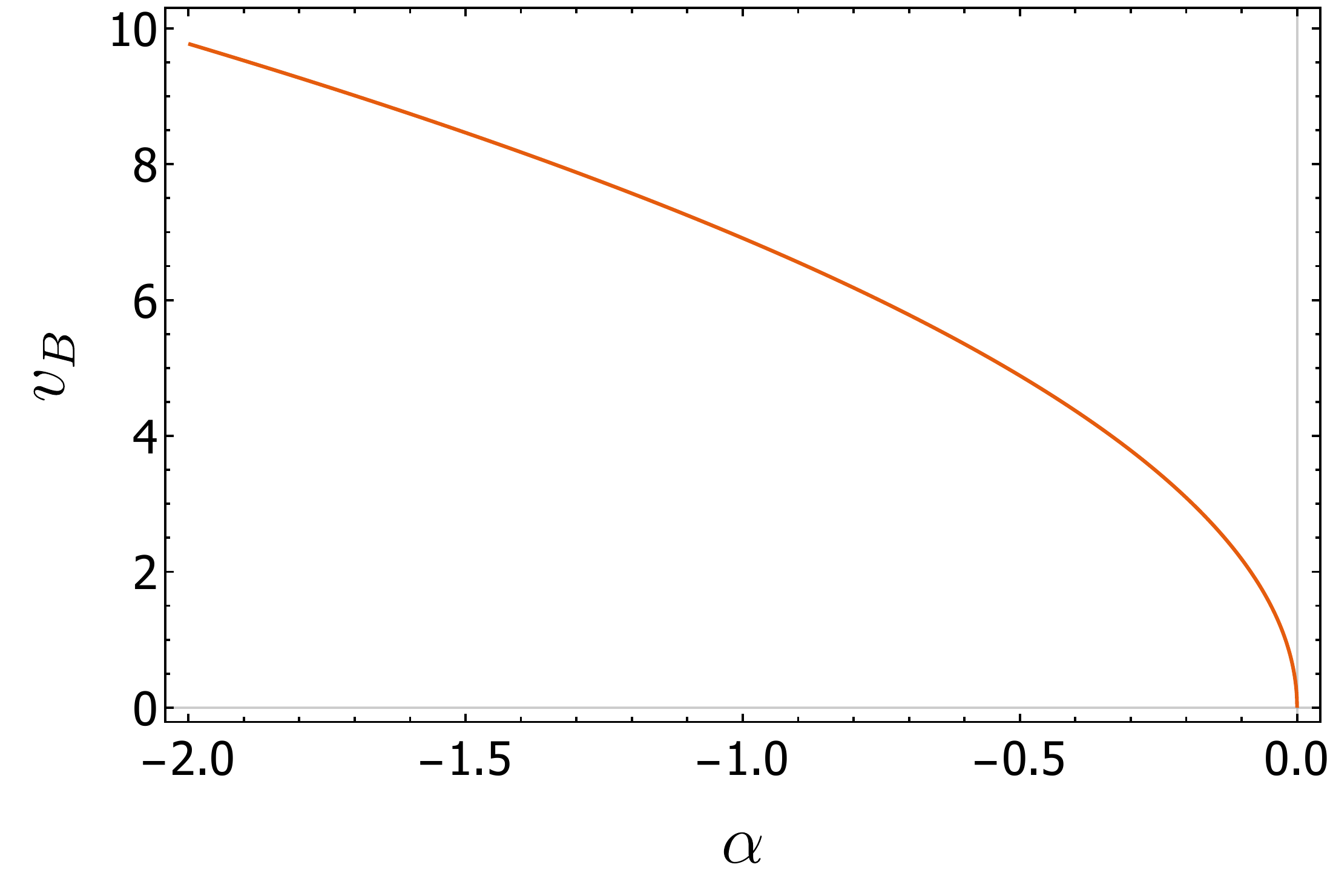}
\end{subfigure}
\caption{Numerical solutions for $\alpha=-5/3$, $\mu_I=\sqrt{3}$, $\mu_A=0.2 \mu_I$ as a function of $t/t_{Pl}$. Dotted and dashed lines represent where bounce and future time singularity happen, respectively. The bounce is centered at the origin of time for convenience, and the values of the parameters are chosen in order to yield graphs that display features in a clear fashion.}
\end{figure}
\twocolumngrid

\subsection{Vacuum case}
As a preliminary case it is useful to consider the vacuum configuration, where $f(v)=1$ and \eqref{hubble function ny} reads after a bit of manipulation as
\begin{equation}\label{hubble function vacuum}
H^2(v)=\frac{\kappa (v^2+\eta_I)\left(P_A(v)\mu_A^2+P_I(v)\mu_I^2\right)}{6v^6\leri{v^2+4\eta_I}^2},
\end{equation}
where
\begin{align}
    P_A(v)=&2v^6+6\eta_I v^4+6\eta_I^2 v^2+2\eta_I^3,\\
    P_I(v)=&2v^4(v^2+2\eta_I)
\end{align}
and $\eta_I\equiv6\alpha\kappa\mu_I^2$.
By inspection of \eqref{hubble function vacuum} we immediately see that for $\alpha>0$ (i.e. $\eta_I>0$) the r.h.s is always positive. This implies that the volume $v$ can span all the positive values, i.e. $v\in \mathbb{R}^+$, and the dynamics is still singular in $v=0$. Big-bounce or turning points are instead related to those values of $v$ where $H^2=0$, corresponding to the zeros of the numerator on the r.h.s. of \eqref{hubble function vacuum}. In particular, in order to distinguish between big-bounce and turning points, we have to select those intervals where $H^2>0$ holds: lower bounds can be identified with big-bounce points and upper bounds with turning points. For $\alpha<0$, therefore, we have to solve the inequality
\begin{equation}
(v^2+\eta_I)\left(P_A(v)\mu_A^2+P_I(v)\mu_I^2\right)\ge 0,
    \label{numerator}
\end{equation}
which a bit of algebraic manipulations reveal to hold in
\begin{equation}
\begin{split}
    &0<v^2<v_T^2\equiv-\eta,\\ 
    &v_B^2<v^2,
\end{split}
\end{equation}
where $v_B^2$ is the only real root of the third order equation in $x=v^2$ which appears in \eqref{numerator} and for whose quite cumbersome expression we address the reader to Appendix ~\ref{appendix A}. We have, therefore, two disconnected domains describing respectively a closed Universe, where singularity is not removed and General Relativity limit cannot be reached ($\phi\rightarrow \infty$ for $v\rightarrow -\eta$), and an open Universe where singularity is classically tamed by a big-bounce in $v=v_B$ and $\phi\rightarrow 1$ for $v\rightarrow +\infty$. In the same limit, moreover, we stress that by virtue of \eqref{solution immirzi DI hy} the Immirzi field boils down to a constant, correctly reproducing the ordinary LQG picture, and that \eqref{hubble function vacuum} can be always recast as the Friedmann equation of a FRW flat Universe filled with a scalar field, i.e.
\begin{equation}
    H^2\sim\frac{\kappa\leri{\mu_I^2+\mu_{A}^2}}{3v^2},
\end{equation}
where with respect to \cite{BombacignoFlavioandMontani2019} (see equations (53) and (54) therein) also the anisotropy energy density concurs in defining the effective energy density for the scalar field. These preliminary results are confirmed by numerical investigations. The big-bounce can be appreciated in Fig.~\ref{fig: PlotV}, obtained integrating equation \eqref{hubble function vacuum} for $v(t)$, after having rescaled all dimensional quantities by the appropriate power of the Planck time $t_{Pl}$ (for the sake of clarity we use the same symbols also for rescaled dimensionless quantities). We see that the volume undergoes a future finite-time {\it singularity} (see \cite{Nojiri:2005sx,Odintsov:2018uaw} for details concerning their classification), corresponding to the pole of equation \eqref{hubble function vacuum} in $v_c=-4\eta_I$, where the Hubble function diverges.
This causes a breakdown of the numerical integration, which we tackle by solving \eqref{hubble function vacuum} separately in the two regions adjacent to the troublesome point and matching the solutions across $v_c=-4\eta_I$. We note, however, that the occurrence of divergences in the derivative of $v$ raises reasonable doubts about whether those solutions can be extended across the singular points without ambiguities, and the viability of such a procedure has to be tested. We refer, in particular, to the geodesic completeness of the solutions and to the behaviour of scalar perturbations, which should be free of pathologies in order to guarantee a physically sensible matching of solutions. These issues will be properly addressed in Sec.~\ref{geodesiccompleteness}, and here we just stress that, in general, creation of particles in the presence of cosmological horizons \cite{Montani:2001fp, Dimopoulos:2018kgl,Ford:1986sy,Contreras:2018two} can lead to additional terms in the Friedman equation, able to stabilize the singular behaviour of the Hubble parameter. 
\\The behaviour of the scalar fields can be analysed via equations \eqref{solution immirzi DI hy} and \eqref{solution phi DI ny} and Fig.~\ref{fig: PhiPsi} shows how the scalaron asymptotically reaches $1$ as $t\to\infty$, while the Immirzi field, as required, relaxes to a constant. Fig.~\ref{fig: BounceAlpha} displays the specific value of the volume at the bounce $v_B$ as a function of the parameter $\alpha$ (see App.~\ref{appendix A} for its explicit formula), while Fig.~\ref{fig: PlotScalefactors} shows the behavior of each scale factor.

\subsection{Radiation and dust}
\begin{figure}
\begin{subfigure}{.5\textwidth}
  \centering
  \caption{\label{fig: AnisotropyRip}Behavior near the finite time singularity (dashed line) for $\mu_R=0$, $\mu_D=0$.}
  \includegraphics[width=1\linewidth]{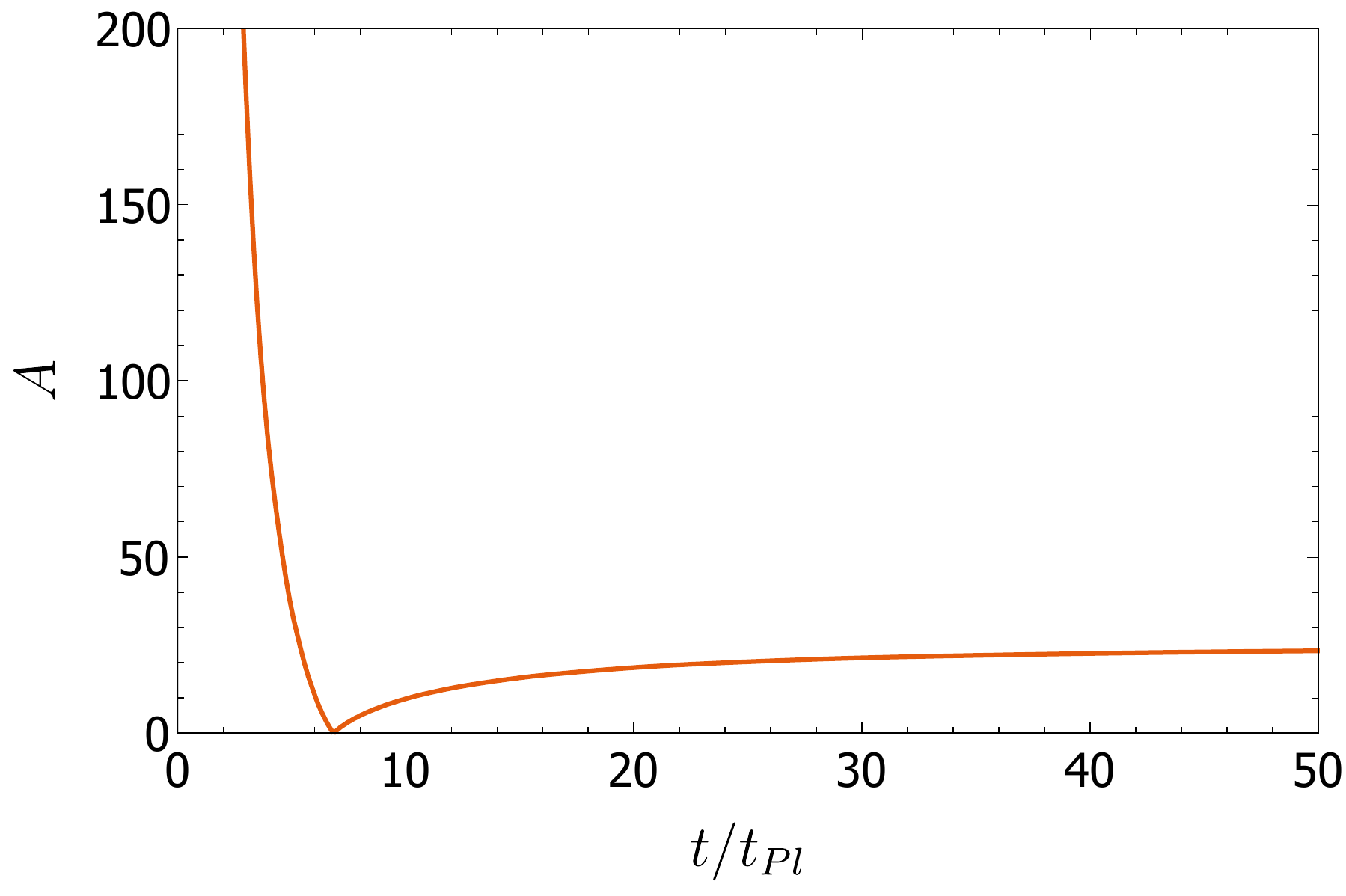}
\end{subfigure}
\newline
\begin{subfigure}{.5\textwidth}
  \centering
  \caption{\label{fig: Anisotropy}Asymptotic behavior for various values of $\mu_R$, $\mu_D$ after the finite time singularity.}
  \includegraphics[width=1\linewidth]{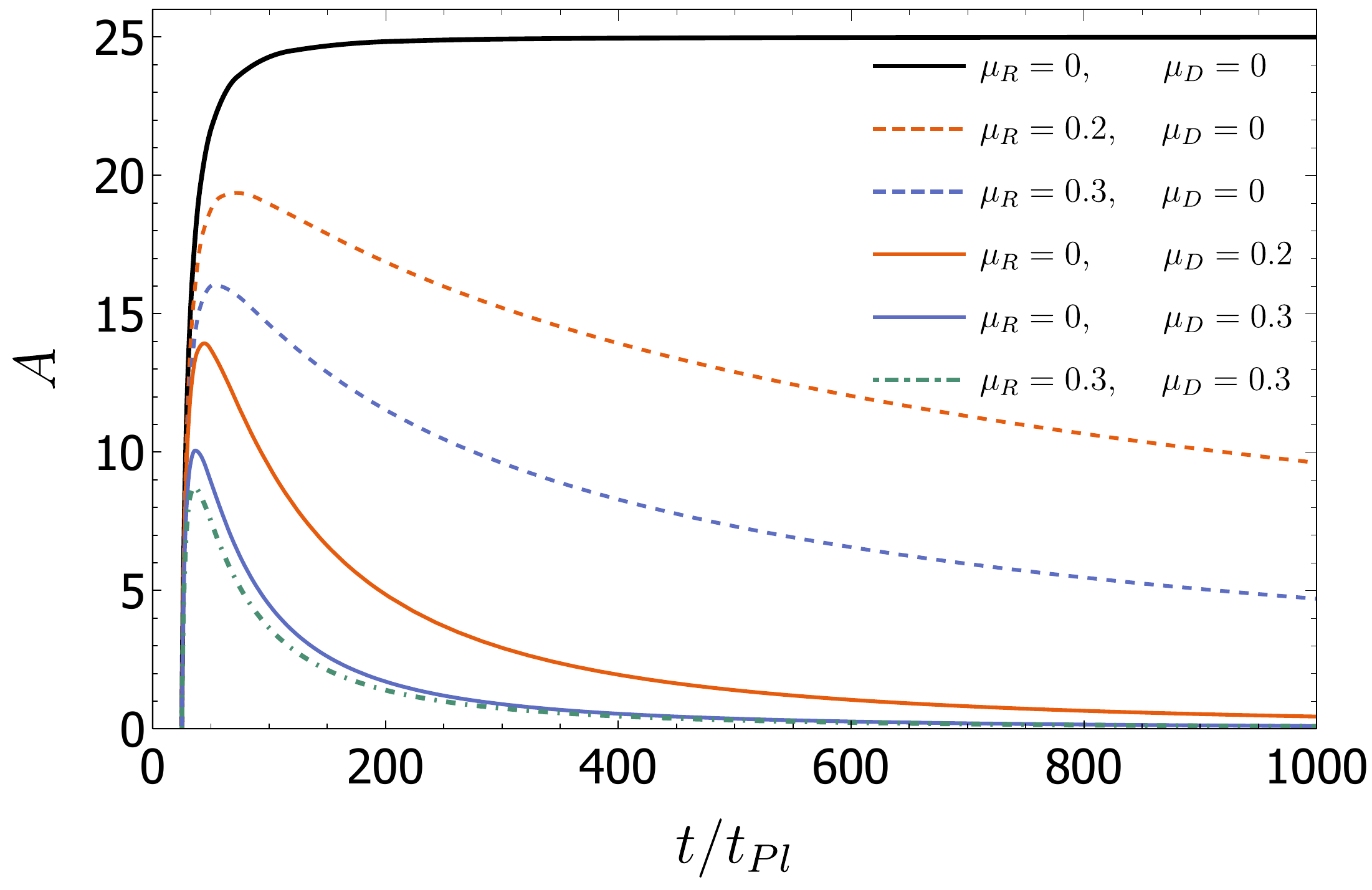}
\end{subfigure}
\caption{Anisotropy degree $A$ as a function of $t/t_{Pl}$ for $\alpha=-5/3$, $\mu_I=\sqrt{3}$, $\mu_A=0.2 \mu_I$.}
\end{figure}
In this section we complete the analysis, including the energy density of radiation and dust (corresponding to $w=1/3$ and $w=0$ in \eqref{energy density}, respectively). In this case \eqref{solution phi DI ny} takes the form  
\begin{equation}
    \phi(v)=\frac{v(v+2\eta_D)}{v^2+\eta_I}
\end{equation}
and we see that the presence of dust introduces an additional zero in $v_P=-2\eta_D\equiv -2\alpha\kappa\mu_D^2$, which as we will discuss below can be lesser, then excluded from the domain of the values of $v$, or greater than the value $v_B$ where the bounce occurs. In the latter case, it affects the evolution of the scale factors $a(t),\,b(t)$ and $c(t)$, since a zero of $\phi$ corresponds to a pole in \eqref{a solved}-\eqref{c solved} and, like for the vacuum, the physical feasibility of such divergences has to be analyzed. 
\\The Hubble rate \eqref{hubble function ny} can be rearranged as
\begin{equation}\label{hubble function matter}
H^2=\frac{\kappa (v^2+\eta_I)\sum_jP_j(v)\mu_j^2}{6v^4\leri{v^3-\eta_D v^2+4\eta_I v + 5\eta_I\eta_D}^2},
\end{equation}
where $j=D,R,A,I$ and
\begin{align}
    P_D(v)=&2v^7+5\eta_D v^6+2(\eta_D^2+\eta_I)v^5+7\eta_I\eta_D v^4\nonumber\\
    &+\frac{7}{2}\eta_I^2 v^3+5\eta_I^2\eta_D v^2,\\
    P_R(v)=&2v^{5/3}(v^5+2\eta_D v^4+2\eta_I v^3+4\eta_I\eta_D v^2\nonumber\\
    &+\eta_I^2 v+2\eta_I^2\eta_D).
\end{align}
In this case the initial singularity is still regularized by a big-bounce but some properties of the solutions are actually different with respect to the vacuum configuration, both in the early phase of the universe and in the late time asymptotic region. Regarding the latter, the behaviour of the scale factors is influenced by both radiation and dust with consequences on the degree of anisotropy of the universe, quantified by the function
\begin{equation}
    A(t) = \frac{\left( H_A^2 + H_B^2 + H_C^2 \right)}{3 H^2} -1.
\end{equation}
For its computation the time evolution of each scale factor is obtained integrating equations \eqref{a solved}, \eqref{b solved} and \eqref{c solved}, once $v(t)$ and $\phi(t)$ are known. While in vacuum $A(t)$ relaxes to a non-vanishing constant at infinity, the presence of matter is able to flatten the curve, providing the isotropization of the universe, as shown in Fig.~\ref{fig: Anisotropy}.

\begin{figure}
\begin{subfigure}{.5\textwidth}
  \centering
  \caption{\label{fig: PlotVPhiNoRip}Volume normalised to $v_B$, $\phi$ and derivative of the Immirzi field.}
  \includegraphics[width=1\linewidth]{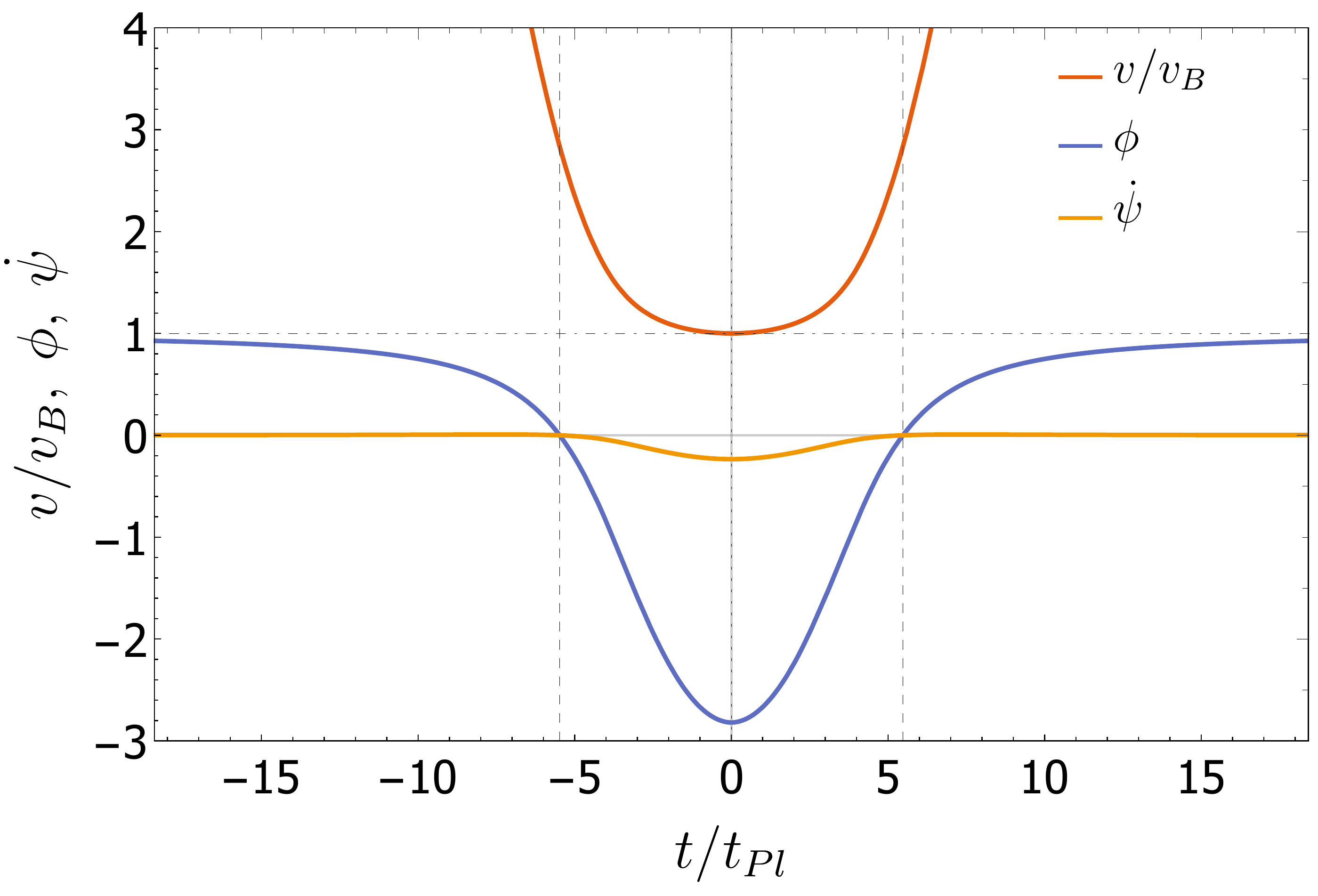}
\end{subfigure}
\newline
\begin{subfigure}{.5\textwidth}
  \centering
  \caption{\label{fig: PlotScaleFactorsNoRip}Scale factors.}
  \includegraphics[width=1\linewidth]{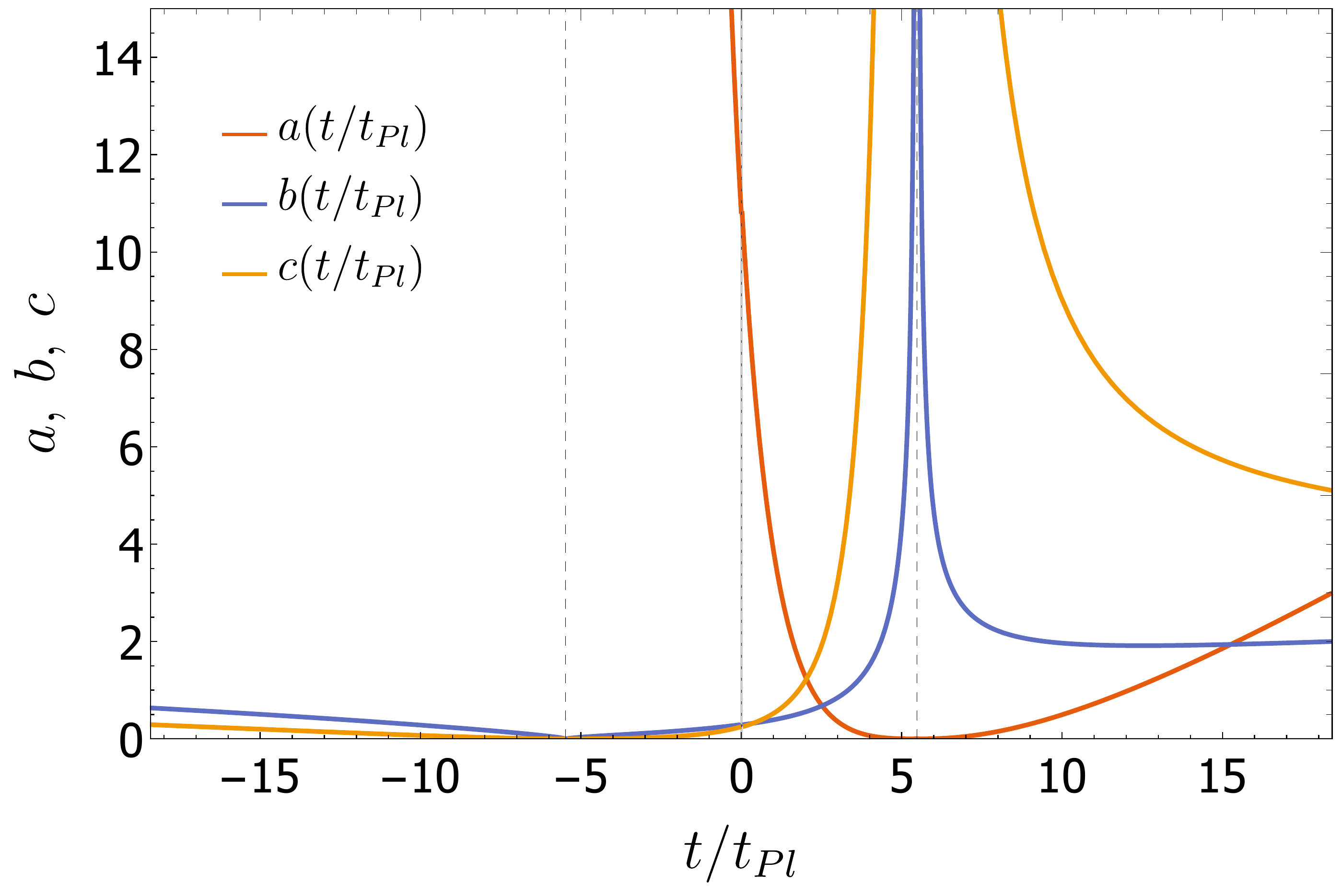}
\end{subfigure}
\caption{\label{fig: norip} Numerical solutions as a function of $t/t_{Pl}$ for $\mu_I=0.057$, $\mu_A=2.4$, $\mu_D=0.365$, $\mu_R=1.56$ and $\alpha=-8.42<\bar{\alpha}$. The dashed lines represent where the scalaron vanishes (color online).}
\end{figure}

Concerning the early phase of the universe, instead, two different scenarios may occur, depending on the value of the parameter $\alpha$. Whenever $\bar{\alpha}<\alpha<0$, where
\begin{equation}\label{alpha bar}
\bar{\alpha} = -2 \mu_I^2 / \mu_D^4,    
\end{equation}
the early behaviour of the volume and the scalar fields is not much altered with respect to the vacuum configuration and the results are similar to those  discussed in the previous section. In this case, of course, the value $v_B$ where the bounce occurs cannot be determined analytically, as it generally depends also on the dust and radiation energy densities. On the other hand, the finite time singularity corresponds now to the real root of the cubic equation in the denominator of \eqref{hubble function matter} and for its expression we refer to App~\ref{appendix A}. 
\\If instead $\alpha<\bar{\alpha}$, the properties of the solutions are fundamentally different (See Fig.~\ref{fig: norip}). In particular, there is no future finite-time singularity for $v$, since the value of the volume at the bounce is always greater than the pole of \eqref{hubble function matter}. The zero of the scalaron, however, is now located after the bounce, leading to the appearance of zeros and singularities for the scale factors, which can interestingly combine without introducing singular points for $v$. The derivative of the Immirzi field, instead, is appreciable only near the bounce, where now it is negative, and rapidly approaches zero, denoting again a constant Immirzi parameter.

\section{Physical implications of curvature divergences}\label{geodesiccompleteness}
\begin{figure}
  \centering
  \includegraphics[width=1\linewidth]{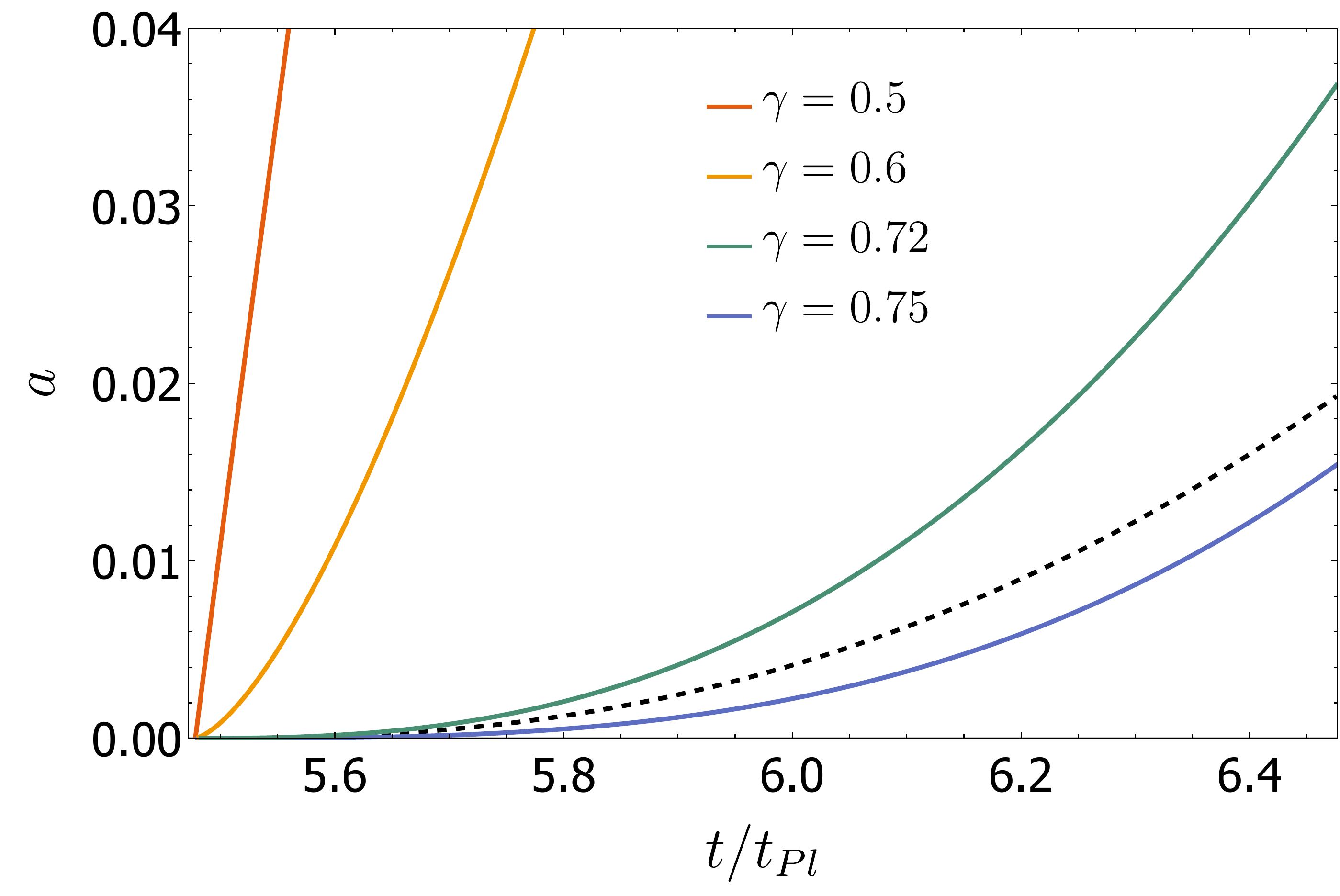}
\caption{\label{fig: geodesics}Outcomes of null geodesics test for $\alpha<\bar{\alpha}$. Scale factor $a(s(t))$ for different values of $\gamma$ and $k_a=C_0=a_0=1$. The dashed-black line represent the numerical solution $a(t)$ reported in Fig. \ref{fig: PlotScaleFactorsNoRip}.}
\end{figure}

A disturbing aspect of the solutions we presented above is the fact that the numerical integration breaks down at a given instant of time, in which the volume of the universe is finite but the Hubble function diverges. Obviously, this divergence in $H$ implies the divergence of various curvature invariants that involve $H$ and its derivatives, which demands a detailed analysis of its physical implications. For concreteness, here we will consider the behavior of geodesics and of scalar perturbations.
\\With elementary algebra, one can show that the geodesic equation for light rays with tangent vector $u^\alpha=dx^\alpha/ds$ leads to \cite{Singh:2011gp,Nomura:2021lzz} 
\begin{eqnarray}
x''&=& -2x' t' H_A, \ \nonumber \\
y''&=& -2y' t' H_B, \ \nonumber \\
z''&=& -2z' t' H_C, \ \nonumber \\
t''&=& -a^2 H_A x'^2-b^2 H_A y'^2-c^2 H_C z'^2,
\end{eqnarray}
where prime denotes derivative with respect to the affine parameter $s$. These equations admit a first integral of the form
\begin{eqnarray}
x'&=& \frac{k_a}{a^2}, \ \nonumber \\
y'&=& \frac{k_b}{b^2},\ \nonumber \\
z'&=& \frac{k_c}{c^2}, \ \nonumber \\
t'&=& \left(\frac{k^2_a}{a^2}+\frac{k^2_b}{b^2} +\frac{k^2_c}{c^2}\right)^{1/2} +C_0,
\label{geodesic derivative}
\end{eqnarray}
with ${k_a,k_b,k_c}$ and $C_0$ representing integration constants. From the basic theory of first-order differential equations, it follows that in those intervals in which the functions $a(t)$, $b(t)$, and $c(t)$ are continuous and non-vanishing, as it occurs for solutions characterized by finite time singularities, both in vacuum and in the presence of matter, the geodesic tangent vector will be unique and well defined. Such cases are clearly non-singular since they are geodesically complete, a result that holds both in the anisotropic and in the isotropic case \cite{Jimenez:2016sgs}. When $\alpha<\bar{\alpha}$, instead, we see that the volume remains finite all over the interval despite the fact that some expansion factors collapse to zero while others diverge (Fig. \ref{fig: PlotScaleFactorsNoRip}). The divergence of the individual expansion factors is not a problem for the geodesics, but the vanishing of some of them may lead to a lack of continuity and, therefore, to the impossibility of a unique extension. To see this, let us consider the situation where one of the scale factors vanishes at some affine parameter $s_c$. In particular, suppose that
\begin{equation}\label{geodesic scale factor}
    a(s)=a_0(s-s_c)^{\gamma},
\end{equation}
with $\gamma>0$. Then, by virtue of \eqref{geodesic derivative}, the relevant equations would be
\begin{align}
    x'&=\frac{k_a}{a_0^2(s-s_c)^{2\gamma}},\\
    t'&= C_0 + \frac{k_a}{a_0(s-s_c)^{\gamma}}.
\end{align}
Now, if we integrate these equations to understand what happens, we obtain:
\begin{align}
    x(s) &= x_c + \frac{k_a(s-s_c)^{1-2\gamma}}{a_0^2(1-2\gamma)},\\
     t(s) &= t_c + C_0(s-s_c) + \frac{k_a(s-s_c)^{1-\gamma}}{a_0(1-\gamma)},\label{geodesic time}
\end{align}
which are smooth if $0<\gamma<1/2$ and $0<\gamma<1$, respectively. Note that if $1/2<\gamma<1$, then $x(s)\xrightarrow[]{s\to s_c}\pm\infty$, which would imply travel to infinity in finite coordinate time. Conversely, if $0<\gamma<1/2$, then the path of a geodesic will cover the range $\{t,x\}\in\; (-\infty,\infty )$, and those geodesics would be complete despite the vanishing of some scale factors at some instant in time. Since this result is strongly dependent on how rapidly the zero is reached, we have to inspect the value of $\gamma$ relative to the solution in Fig.~\ref{fig: PlotScaleFactorsNoRip}. This can be performed in the following way: for various values of $\gamma$, eq. \eqref{geodesic time} can be inverted for $s(t)$ which, substituted in \eqref{geodesic scale factor}, gives the scale factor $a(s(t))$ as a function of $t$. This can be compared with the numerical solution in Fig. \ref{fig: PlotScaleFactorsNoRip} obtained in the previous section. The results shown in Fig.\ref{fig: geodesics} indicate that the solution approaches zero too rapidly, corresponding to a value of $\gamma$ larger than $1/2$. We are thus forced to conclude that the example shown in Fig. \ref{fig: PlotScaleFactorsNoRip} does represent a geodesically incomplete  space-time.

Since geodesics describe the propagation of high-frequency (or infinite frequency) modes, it is convenient also to have a look at the behavior of scalar field perturbations in order to test how finite frequencies evolve upon encountering a divergence in the Hubble function. In this regard, one can consider a generic scalar field or simply assume the existence of small inhomogeneous perturbations of the field $\psi$ around a given homogeneous background solution. In all such cases, for a scalar mode of the form $\sigma_{\vec{k}}(t,\vec{x})=\Theta(t) e^{i \vec{k}\cdot\vec{x}}$, one finds an equation of the form
\begin{equation}\label{eq:A-equation}
\ddot{\Theta}+h(v) \frac{\dot v}{v} \dot \Theta+\left(\frac{k_x^2}{a^2}+\frac{k_y^2}{b^2}+\frac{k_z^2}{c^2}\right)\Theta=0 \ , 
\end{equation}
where $h(v)$ represents some regular function of the volume $v$ and $\vec{k}=(k_x,k_y,k_z)$ represents a set of constants. From this expression, it is evident that scalar modes feel the presence of the individual scale factors $a,b,$ and $c$, and of the Hubble function $3H=\dot{v}/v$.
\\We will now discuss generic situations and will then particularize to the cases found in our model. First of all, we note that if the scalar factors $a,b,$ and $c$ do not vanish anywhere, then the last term in (\ref{eq:A-equation}) is well behaved and bounded. Any potential problems should come from the damping term involving the Hubble function $H=\dot v/3v$, which for finite time singularity diverges. In vacuum, in particular, one finds that as $v^2\to 4|\eta_I|\equiv v_c^2$ the Hubble function can be approximated as 
\begin{equation}
H^2\approx \frac{\kappa \mu_I^2}{2^{10}}\frac{(32+27\lambda^2_{AI})}{(v-v_c)^2}.
\end{equation} 
From this equation we see that the divergent piece $\dot{v}/v$ goes like
\begin{equation}
\frac{\dot{v}}{3v}\approx \pm \sqrt{\frac{\kappa \mu_I^2}{2^{10}}\frac{(32+27\lambda^2_{AI})}{(v-v_c)^2}}\equiv \pm\frac{C_1}{|v-v_c|} \ , 
\end{equation}
where the $\pm$ sign corresponds to the expanding/contracting phase. This result can be used to write
\begin{eqnarray}
\dot{v}&\approx& \pm\frac{3v_c C_1}{|v-v_c|}, \\
|v-v_c|&\approx&  \sqrt{6 v_c C_1} |t-t_c|^{1/2}  \ .
\end{eqnarray}
This solution shows that in the vacuum case scalar perturbations satisfy a second-order linear differential equation with an avoidable singular point at $t=t_c$, where $v=v_c$. The dominant contribution in the neighborhood of $t_c$ can be obtained by neglecting the last term in (\ref{eq:A-equation}), such that we are left with 
\begin{equation}\label{eq:A-eq-approx}
\ddot{\Theta}\pm\frac{\tilde{h}_c}{|t-t_c|^{1/2}} \dot \Theta\approx 0 \ , 
\end{equation}
where $\tilde{h}_c\equiv h(v_c) \sqrt{\frac{3C_1}{2v_c}}$. It leads to 
\begin{equation}
\Theta(t)\approx \Theta_c+\frac{\dot \Theta_c}{2\tilde{h}_c^2}e^{\mp2\tilde{h}_c|t-t_c|^{1/2}}\left(1\pm2\tilde{h}_c|t-t_c|^{1/2}\right) \ ,
\end{equation}
with $\Theta_c, \dot \Theta_c$ integration constants. As expected, this expression is finite regardless of the sign of the parameter $\tilde{h}_c$ and confirms that scalar field perturbations remain bounded around $t_c$ despite the divergence in the Hubble function.

Following a similar reasoning, we can explore what happens to scalar perturbations in scenarios with dust and radiation such as those corresponding to Eq. (\ref{hubble function matter}). In that situation, the worst case scenario (or strongest divergence) would correspond to having a triple root in the denominator, such that $H^2\approx C_2/(v-v_c)^6$. This would lead to $|v-v_c|\sim |t-t_c|^{1/4} $ and $\dot v/v\sim \pm1/|t-t_c|^{3/4}$, from which one finds
\begin{equation}
    \ddot{\Theta}\pm\frac{\tilde{h}_c}{|t-t_c|^{3/4}}\dot{\Theta}\approx 0,
\end{equation}
where now $\tilde{h}_c=h(v_c)(3 C_2/v_c^3)^{1/4}$. Therefore, one has
\begin{align}
    \Theta \approx & \Theta_c- \frac{\dot{\Theta}_c}{32 \tilde{h}_c^4} e^{\mp4\tilde{h}_c|t-t_c|^{1/4}} \left(  3\pm12\tilde{h}_c|t-t_c|^{1/4} \right.\nonumber\\
    &\left.+24\tilde{h}_c^2 |t-t_c|^{1/2} \pm 32\tilde{h}_c^3 |t-t_c|^{3/4} \right),
\end{align}
which is easy to see to be again bounded. We conclude, therefore, that in the nearby of finite time singularities, both in vacuum and in the presence of matter, scalar perturbations are always well behaved, guaranteeing together with geodesic completeness the physical feasibility of such solutions. Conversely, when the Hubble function is regular but one of the scale factors vanishes, equation \eqref{eq:A-equation} describes an harmonic oscillator with a time dependent frequency, 
\begin{equation}
\ddot{\Theta}(t)+\frac{k_x^2}{a^2(t)}\Theta(t)\approx0,
\end{equation}
which diverges as $a(t)\to 0$. Though there might be cases in which the integrated solution yields a finite result, for the values of $a(t)$ obtained numerically in the previous section neither geodesics nor scalar perturbations are well behaved.

\section{Conclusion}\label{conclusions}
In this paper, we proposed an extension of the Nieh-Yan form to the framework of metric-affine gravity, by including an additional term depending on nonmetricity and featuring two parameters ($\lambda_1$, $\lambda_2$), which allow to restore the projective invariance and the topological character. In particular, we showed that projective invariance is a property which can be independently recovered by setting the values of the parameter as $\lambda_1=\lambda_2=\lambda$, whereas topologicity is only obtained for $\lambda=1$.
\\ We considered, then, a model described by the Lagrangian $\sqrt{-g}F(R,NY_{gen})$, which conveniently expressed in the Jordan frame features two new scalar fields. We identify these additional scalar degrees as a $f(R)$-like scalaron $\phi$ and the Immirzi field $\beta$. In this framework, the latter acquires dynamical character and a potential term in a more natural way than in previous treatments, where these features were introduced by hand in the action.
\\Two different effective scalar tensor theories arise, depending on the values of $\lambda_1$ and $\lambda_2$. If they coincide, i.e. in the projective invariant case, the theory is endowed with an additional dynamical degree of freedom, the Immirzi field, while the scalaron is algebraically related to the latter via a modified structural equation. Models with $\lambda_1\neq\lambda_2$, instead, are non-dynamical, in the sense that both can be expressed as a function of the trace of the stress energy tensor, by analogy with Palatini $f(R)$ theories. In particular, this implies that in vacuum both scalar fields boil down to constant values $\phi_0$ and $\beta_0$, and we recover GR with a cosmological constant which now depends on the value of the Immirzi parameter via the potential term, i.e. $\Lambda=W(\phi(\beta_0),\beta_0)/2$.
\\ Eventually, we controlled that in order to reproduce previous results in the literature where nonmetricity is a priori neglected, the vanishing of the latter must be enforced as a constraint in the action via a Lagrange multiplier. This comparison with previous works yields two relevant outcomes. On the one hand, the reduction to the correct Einstein-Cartan version of the NY term is a consistency check on the specific expression for $NY_{gen}$ we defined in \eqref{NY general}. On the other hand, the results obtained in the analysis explain why, despite the violation of projective symmetry, the degrees of freedom of the corresponding theories are healthy (as was the case for the models previously analysed in \cite{Calcagni2009,Mercuri2006,Mercuri2008,Mercuri2009,Mercuri2009a,Bombacigno2016,Bombacigno2018,Bombacigno2019}). This is due to an on-shell recovery of projective invariance assured by the condition $S_\mu=0$ (see \eqref{NYgen components}).
\\ We considered, thus, in more detail the dynamical models. We first established an equivalence with the subclass of DHOST theories which are experimentally compatible and verified that, in general, the Immirzi field is always devoid of ghost instabilities.
\\ Then, we specialized to a quadratic model described by $F(R,NY_{gen})=R+\alpha R^2+ NY_{gen}$ and we looked for cosmological solutions starting from a Bianchi I metric. For negative values of the parameter $\alpha$, we found solutions characterized by different behaviors of the fields. A common feature of these solutions is that the big bang singularity is removed in favour of a big bounce scenario, in which the volume like variable undergoes a contraction up to a minimum value and then bounces back, re-expanding symmetrically in another branch. This behavior also arises in isotropic scenarios when the $NY_{gen}$ term is not included \cite{Barragan:2009sq,Barragan:2010uj}, though in Bianchi I configurations, those models fail to generate non-singular solutions \cite{Barragan:2010qb}. 
A first investigation carried out in absence of any matter content revealed the presence of a finite time future singularity after the bounce, in which the first derivative of the volume becomes infinite, while the volume itself and the scale factors are finite. Nonetheless, in the neighborhood of this point, null geodesics are well behaved and scalar perturbations bounded, which allows us to conclude that the solution is physically acceptable. The scalaron and the Immirzi field reach a maximum during the bounce and relax to constant values at later times, where the standard LQG picture, with $\phi=1$ and a constant Immirzi parameter $\beta=\beta_0$, is recovered.
When dust and radiation are included in the analysis, the solutions separate into two classes, depending on the value of $\alpha$ compared to $\bar{\alpha}$, defined in \eqref{alpha bar}. If $\alpha<\bar{\alpha}$, the only difference with respect to the vacuum configuration is that both dust and radiation are able to provide an isotropization effect at late times, a feature that is absent in vacuum. In the range $\bar{\alpha}<\alpha<0$, instead, we find a different scenario. The future finite time singularity in the evolution of the volume of the universe  is absent, and the latter and its derivatives are always regular, but the singularity is then transposed to the evolution of the scale factors. These encounter either a zero or a singularity at finite time $t_c$ after the bounce. Concurrently, the scalar field $\phi$ vanishes, reaching negative values in the proximity of the bounce, while the Immirzi field continues to relax to a constant for late times. In this case, however, the study of null geodesics shows that they cannot be extended across the singular point, where also scalar perturbations grow in time, leading us to regard such solutions as unphysical.
\\ Summarizing, we showed that a Bianchi I cosmology can be characterized, in the present geometrical framework, by a big-bounce scenario for the early universe and an isotropization behavior for the late universe. Furthermore, the typical singularities appearing in this type of geometrical Lagrangian, when applied to a cosmological sector, say the various versions of the so-called
Big-Rip \cite{Nojiri:2005sx,Odintsov:2018uaw}, are here associated to a viable phenomenology.
\\The value of having investigated the Bianchi I model, moreover, consists in the general role that a Kasner-like dynamics plays in constructing the general cosmological solution \cite{doi:10.1080/00018738200101428,Montani:2011zz,doi:10.1080/00018736300101283} (see also \cite{Antonini:2018gdd,Giovannetti:2021bqh}) for a Bianchi I bouncing cosmology in the polymer quantization scheme.
\\ We conclude by observing that, in the considered scenario, the Immirzi field always approaches, in the late universe, a constant value, according to the idea that a geometrical Lagrangian compatible with LQG can be recovered as a result of the cosmological dynamics, from more general affine formulations of the geometrodynamics. 

\acknowledgments
The work of F. B. is supported by the Fondazione Angelo della Riccia grant. This work is supported by the Spanish project  FIS2017-84440-C2-1-P (MINECO/FEDER, EU), the project PROMETEO/2020/079 (Generalitat Valenciana), the project i-COOPB20462 (CSIC) and the Edital 006/2018 PRONEX (FAPESQ-PB/CNPQ, Brazil, Grant 0015/2019). This article is based upon work from COST Action CA18108, supported by COST (European Cooperation in Science and Technology).

\appendix
\section{}\label{appendix A}
In this appendix we report the expression for the volume at the bounce ($v_B$) in the vacuum case, in terms of the Immirzi and anisotropy energy densities, i.e.
\begin{equation}
\begin{split}
    v_B^2=-&\frac{1}{3(1+\lambda_{AI}^2)\eta_I}\left[\frac{2^{1/3}(4+3\lambda_{AI}^2)}{Q_{\frac{4}{3}}(\lambda_{AI})}\right.\\
    &\left.+(2+3\lambda_{AI}^2)\eta_I^2+\frac{Q_{\frac{4}{3}}(\lambda_{AI})}{2^{1/3}}\eta_I^4\right],
\end{split}
\end{equation}
where we define the quantities
\begin{align}
    &Q_{\frac{4}{3}}(\lambda_{AI})\equiv\sqrt[3]{16-Q_2(\lambda_{AI})\lambda_{AI}+45\lambda_{AI}^2+27\lambda_{AI}^4}\\
    &Q_2(\lambda_{AI})\equiv3\sqrt{3}\sqrt{32+91\lambda_{AI}^2+86\lambda_{AI}^3+27\lambda_{AI}^4},
\end{align}
and the value of the volume at the finite time singularity in the presence of matter, which results to depend on the Immirzi and the dust energy densities, i.e.
\begin{equation}
v_r=\frac{1}{3}\leri{\eta_D-\frac{2^{1/3}(12\eta_I-\eta_D^2)}{P_1(\eta_I,\eta_D)}+\frac{P_1(\eta_I,\eta_D)}{2^{1/3}}},
\end{equation}
where we introduced
\begin{align}
    &P_{1}(\eta_I,\eta_D)\equiv\leri{2\eta_D^3-17\eta_D\eta_I+P_{\frac{5}{3}}(\eta_I,\eta_D)}^{5/3}\\
    &P_{\frac{5}{3}}(\eta_I,\eta_D)\equiv 48\sqrt{3}\sqrt{\eta_I (\eta_I+4\eta_D^2)\leri{\eta_I-\frac{5\eta_D^2}{256}}}.
\end{align}

\bibliography{references}{}

\end{document}